\documentclass[aps,prd,preprint,groupedaddress,showpacs]{revtex4}
\usepackage{amssymb,amsmath,graphicx}

\begin{document}
\title{Stochastic Lorentz forces on a point charge moving near the conducting plate}
\author{Jen-Tsung Hsiang}
\email{cosmology@mail.ndhu.edu.tw}
\author{Tai-Hung Wu}
\author{Da-Shin Lee}
\email{dslee@mail.ndhu.edu.tw}\thanks{corresponding author.}
\affiliation{Department of Physics, National Dong Hwa University,
Hualien, Taiwan, R.O.C.}

\date{\today}

\begin{abstract}
The influence of quantized electromagnetic fields on a
nonrelativistic charged particle moving near a conducting plate is
studied. We give a field-theoretic derivation of the nonlinear,
non-Markovian Langevin equation of the particle by the
method of Feynman-Vernon influence functional. This stochastic
approach incorporates not only the stochastic
noise manifested from electromagnetic vacuum fluctuations, but also dissipation backreaction on a charge
in the form of the retarded Lorentz forces. Since
the imposition of the boundary is expected to anisotropically modify
the effects of the fields on the evolution of the particle, we
consider the motion of a charge undergoing small-amplitude
oscillations in the direction either parallel or normal to the plane
boundary. Under the dipole approximation for nonrelativistic
motion, velocity fluctuations of the charge are found to grow
linearly with time in the early stage of the evolution at the rather different rate, revealing strong anisotropic behavior. They are then asymptotically saturated as a result
of the fluctuation-dissipation relation, and the same saturated value is found for the motion in both directions. The observational consequences are discussed.
\end{abstract}

\maketitle

\allowdisplaybreaks

\section{Introduction}
The manifestation of vacuum fluctuations can be visualized through a
mechanical effect on a materialized body. One of the most celebrated
examples is the attractive Casimir force between two parallel
conducting plates~\cite{CA1}. This phenomenon in general can be characterized by fluctuations under geometric constraints such that its spectrum on long wavelength
modes is modified. Nevertheless, this induced-force effect can also
be probed through the coupling to a test particle. For example, consider an atom
in its ground state  as a test particle located near a perfectly
conducting plate. The atom then experiences a position-dependent
energy shift due to the boundary effect on vacuum fluctuations from
which to give rise to an attractive Casimir-Polder force toward the
plate~\cite{CA2}. Thus, the presence of the boundary is expected to
anisotropically change vacuum fluctuations. In this paper a charged test particle serves as a probe to understand the nature of electromagnetic vacuum fluctuations by observing their effects on the test-particle's trajectory.

When a charged particle interacts with quantized electromagnetic
fields, not only do the expectation values of fields determine the
mean trajectory of the charge, but the accompanying
quantum fluctuations also drive the charge into a
zig-zag motion. The dynamics of the particle and field interaction has
been studied quantum-mechanically in the system-plus-environment
approach~\cite{CAL}. We treat the particle as the system of
interest, and the degrees of freedom of fields as the environment.
The influence of fields on the particle can be obtained with the
method of Feynman-Vernon influence functional, by integrating out
field variables within the context of the closed-time-path formalism
~\cite{FE,SC}. The Langevin equation can then be derived
by ignoring the intrinsic quantum uncertainty of the particle, which
is assumed to be much smaller than the resolution of the position
measurements. This stochastic approach incorporates both
dissipative backreaction arising from the interaction with fields, and a stochastic noise owing to the quantum fluctuations of fields.
In particular, the non-uniformity of the charge's motion will result
in radiation that backreacts on itself through the electromagnetic
self-force. The stochastic noise, which encodes the influence of
quantum statistics of  fields, drives the charge into a fluctuating
motion~\cite{HU1,JA}. Furthermore, the noise-averaged result reduces to the known
Abraham-Lorentz-Dirac equation with the self-force given by a
third-order time derivative of the position as expected~\cite{HU1}.

The anisotropy of electromagnetic vacuum fluctuations in the
presence of the conducting plate has been studied via an
interference experiment of the electrons, and is manifested in the
form of the amplitude change and phase shift of the interference
fringes~\cite{BA,FO,MA,JT}. In the previous article~\cite{JT}, we
employ the method of influence functional, and obtain the
evolution of the reduced density matrix of the electron with
self-consistent backreactions from quantized electromagnetic fields.
Under the classical approximation with prescribed electron's
trajectories, it is shown that the modulus of the exponent in the
influence functional describes the change of the interference
contrast, and its phase  results in an overall shift of the
interference pattern. It is also found that the presence of the
boundary anisotropically modifies the contrast of the interference
fringes. In Ref.~\cite{YU}, the Brownian motion of a charged
particle coupled to electromagnetic vacuum fluctuations near a
perfectly conducting plate is studied for the case that the particle
barely moves; thus its dissipation effects is ignored. The behaviors of velocity
fluctuations are shown different for the particle's motion in the
directions perpendicular and parallel to the boundary plane.  In
this paper, we wish to further explore the anisotropic nature of
 vacuum fluctuations due to the boundary by the nontrivial motion of the charged particle where dissipative backreaction is incorporated in a
consistent manner. In the presence of the boundary, we expect that
radiation emitted by the charge in nonuniform motion should be
bounced back and then impinge upon the original charge at later times. This
will give rise to an additional retardation effect, which in turn
results in a non-Markovian evolution of the particle. Thus, its mean
trajectory will be altered. Here we will apply red the approach of influence functional from which to derive the Langevin equation beyond the mean-field
approximation. The trajectory fluctuations off the mean value driven
by the stochastic noise will be studied with dissipation backreaction taken into account in a way that a underlying fluctuation-dissipation relation is obeyed.

Our presentation is organized as follows. In Sec.~\ref{sec1}, we
introduce the closed-time-path formalism to describe the evolution
of the density matrix of a nonrelativistic charge coupled to
quantized electromagnetic fields. We trace out field variables to
obtain the coarse-grained effective action from which the Langevin equation is derived. The resulting stochastic Lorentz force, which can be cast into a gauge invariant expression
will be discussed in Sec.~\ref{sec2}. The solutions to the Langevin
equation under a dipole approximation can be found through the
Laplace transform for the charge's motion either parallel or perpendicular
to the boundary plane in Sec.~\ref{sec4}. Thus,  velocity
fluctuations are obtained in Sec.~\ref{sec5}. The results are
summarized and discussed in Sec.~\ref{sec6}.

The Lorentz-Heaviside units with $\hbar=c=1$ will be adopted unless
otherwise noted. The metric is $\eta^{\mu\nu}=\text{diag}(+1,-1,-1,-1)$.

\section{influence functional and Langevin equation}\label{sec1}
We consider the dynamics of a nonrelativistic particle of
charge $e$ interacting with quantized electromagnetic fields. In the Coulomb gauge, $\mathbf{\nabla}\cdot \mathbf{A}=0$, the Lagrangian is expressed as
\begin{align}\label{lagrangian-charge}
    L[\mathbf{q},\mathbf{A}_{\mathrm{T}}]=\frac{1}{2}m\dot{\mathbf{q}}^2-V(\mathbf{q})-&\frac{1}{2}
    \int\!d^3\mathbf{x}\,d^3\mathbf{y}\;\varrho(x;\mathbf{q})G(\mathbf{x},\mathbf{y})\varrho(y;\mathbf{q})\notag\\
    &\qquad\qquad\qquad+\int\!d^3\mathbf{x}\;\left[\frac{1}{2}(\partial_\mu\mathbf{A}_{\mathrm{T}})^2+\mathbf{j}\cdot\mathbf{A}_{\mathrm{T}}\right]\,,
\end{align}
in terms of the transverse components of the gauge potential $\mathbf{A}_{\mathrm{T}}$, and the position $\mathbf{q}$ of the charged particle. The instantaneous Coulomb Green's function $G(\mathbf{x},\mathbf{y})$ satisfies the Gauss's law. The charge and current densities take the form, respectively,
\begin{equation}
\varrho (x; {\bf q}(t)) =e\, \delta^{(3)} ( {\bf x}-{\bf q}(t) ) \,,\qquad\qquad
{\bf j} (x;  {\bf q}(t)) = e\,  \dot{\bf q} (t) \, \delta^{(3)} ( {\bf
x}-{\bf q}(t) ) \, . \label{charge-current}
\end{equation}
Let ${\hat \rho}(t)$ be the density matrix of the particle-field
system, and then it evolves unitarily according to
\begin{equation}
{\hat \rho} (t_f) = U(t_f, t_i) \, {\hat \rho} (t_i) \, U^{-1} (t_f,
t_i )
\end{equation}
with $ U(t_f,t_i) $ the time evolution operator of the total system. The nonequilibrium
partition function can be defined by taking the trace of the density
matrix over the particle and field variables,
\begin{equation}
{\cal Z} =  {\rm Tr} \left\{ U(t_f,t_i) \, {\hat \rho} (t_i) \,
U^{-1} (t_f,t_i) \right\} \, .
\end{equation}
It is convenient to assume that the state of the particle-field at
an initial time $t_i$ is factorizable as ${\hat \rho} (t_i) = {\hat
\rho}_{e} (t_i) \otimes {\hat \rho}_{{\bf A}_{\rm T}} (t_i)$. The
more sophisticated scheme of the density matrix involving initial
correlations can be found in Ref.~\cite{GR}. We also assume that the
particle initially is in a localized state, and thus its density matrix can be
expanded by the position eigenstate of the eigenvalue
$\mathbf{q}_i$,
\begin{equation}
{\hat \rho}_{e} (t_i) = \left|\mathbf{q}_i,t_i \right>\left<
\mathbf{q}_i, t_i \right|\,. \label{initialcondition_e}
\end{equation}
The electromagnetic field at the time $t_i$ is assumed in thermal equilibrium at
temperature $T=1/\beta$, and thus its density operator takes the
form
\begin{equation}
\hat{\rho}_{{\bf A}_{\rm T}} (t_i) =e^{-\beta  H_{{\bf A}_{\rm T}}}/\operatorname{Tr}\left\{e^{-\beta  H_{{\bf A}_{\rm T}}}\right\}\,, \label{initialcondphi}
\end{equation}
where $ H_{{\bf A}_{\rm T}} $ is the Hamiltonian of free vector
potentials in the Coulomb gauge. Later, we will focus on the case that the initial
state of fields is the vacuum state in the zero-temperature limit $\beta\rightarrow \infty$. The nonequilibrium partition functional can be
computed with the help of the path integral along the contour in the
complex time plane by taking the limits, $ t_i \rightarrow -\infty$
and $ t_f \rightarrow +\infty$, and it is given by~\cite{JT},
\begin{equation}
     {\cal Z}=\int\!d^3\mathbf{q}_f \;\int^{\mathbf{q}_f}_{\mathbf{q}_i}\!\!\mathcal{D}\mathbf{q}^+\!\!\int^{{\mathbf{q}}_f}_{\mathbf{q}_i}\!\!\mathcal{D}\mathbf{q}^-\;\exp\left[\frac{i}{\hbar}
\int_{-\infty}^{\infty}dt\left(L_{e}[\mathbf{q}^+]-L_{e}[\mathbf{q}^-]\right)\right]\mathcal{F}[\mathbf{j}^+,\mathbf{j}^-]\,,
\end{equation}
where the Lagrangian $L_e[\mathbf{q}]$ is
\begin{equation}\label{lag}
    L_e\bigl[\mathbf{q}\bigr]=\frac{1}{2}m\dot{\mathbf{q}}^2-V(\mathbf{q})-\frac{1}{2}\int\!d^3\mathbf{x}\,d^3\mathbf{y}\;\varrho(x;\mathbf{q})\,G(\mathbf{x},\mathbf{y})\,\varrho(y;\mathbf{q})\,.
\end{equation}
The influence functional
$\mathcal{F}\left[\mathbf{j}^{+},\mathbf{j}^{-}\right]$, after
tracing out field variables,
 can be written in terms of real-time Green's functions of  vector potentials,
\begin{align}
    \mathcal{F}\left[\mathbf{j}^{+},\mathbf{j}^{-}\right]=\exp\biggl\{-\frac{1}{2\hbar^2}\int d^4x\!\!\int\!d^4x'\Bigl[&j^+_i(x;\mathbf{q}^+(t))\,\bigl<A^{+i}_{\mathrm{T}}(x)A^{+j}_{\mathrm{T}}(x')\bigr>\,j^+_j(x';\mathbf{q}^+(t'))\Bigr.\biggr.\notag\\
    -\;&j^+_i(x;\mathbf{q}^+(t))\,\bigl<A^{+i}_{\mathrm{T}}(x)A^{-j}_{\mathrm{T}}(x')\bigr>\,j^-_j(x';\mathbf{q}^-(t'))\notag\\
    -\;&j^-_i(x;\mathbf{q}^-(t))\,\bigl<A^{-i}_{\mathrm{T}}(x)A^{+j}_{\mathrm{T}}(x')\bigr>\,j^+_j(x';\mathbf{q}^+(t'))\notag\\
    \biggl.\Bigl.+\;&j^-_i(x;\mathbf{q}^-(t))\,\bigl<A^{-i}_{\mathrm{T}}(x)A^{-j}_{\mathrm{T}}(x')\bigr>\,j^-_j(x';\mathbf{q}^-(t'))\Bigr]\biggr\}\,, \label{influencefun}
\end{align}
and contains full information about the influence of quantized
electromagnetic fields. Here the
explicit $\hbar$ dependence is restored in the expressions. The
Green's functions  are defined by
\begin{eqnarray*}
    \bigl<A^{+i}_{\mathrm{T}}(x)A^{+j}_{\mathrm{T}}(x')\bigr>&=&\bigl<A_{\mathrm{T}}^i(x)A_{\mathrm{T}}^j(x')\bigr>\,\theta(t-t')+\bigl<A_{\mathrm{T}}^j(x')A_{\mathrm{T}}^i(x)\bigr>\,\theta(t'-t)\,,\nonumber\\
    \bigl<A^{-i}_{\mathrm{T}}(x)A^{-j}_{\mathrm{T}}(x')\bigr>&=&\bigl<A_{\mathrm{T}}^j(x')A_{\mathrm{T}}^i(x)\bigr>\,\theta(t-t')+\bigl<A_{\mathrm{T}}^i(x)A_{\mathrm{T}}^j(x')\bigr>\,\theta(t'-t)\,,\nonumber\\
    \bigl<A^{+i}_{\mathrm{T}}(x)A^{-j}_{\mathrm{T}}(x')\bigr>&=&\bigl<A_{\mathrm{T}}^j(x')A_{\mathrm{T}}^i(x)\bigr>\equiv\mathrm{Tr}\left\{\rho_{\mathbf{A}_{\mathrm{T}}}\,A_{\mathrm{T}}^j(x')A_{\mathrm{T}}^i(x)\right\}\,,\nonumber \\
    \bigl<A^{-i}_{\mathrm{T}}(x)A^{+j}_{\mathrm{T}}(x')\bigr>&=&\bigr<A_{\mathrm{T}}^i(x)A_{\mathrm{T}}^j(x')\bigr>\equiv\mathrm{Tr}\left\{\rho_{\mathbf{A}_{\mathrm{T}}}\,A_{\mathrm{T}}^i(x)A_{\mathrm{T}}^j(x')\right\}\,.\label{noneqgreenfun}
\end{eqnarray*}
It is found more convenient to change the variables $\mathbf{q}^{+}$ and $\mathbf{q}^{-}$
to the average and relative coordinates,
\begin{equation}
    \mathbf{q}=\frac{1}{2}\left(\mathbf{q}^+ +\mathbf{q}^-
\right)\,,\qquad\mathbf{r}=\mathbf{q}^+ -\mathbf{q}^- \,.
\end{equation}
The nonequilibrium partition function in terms of the coarse-grained action becomes
\begin{equation}
    \mathcal{Z}=\int d\mathbf{q}_f\!\int\mathcal{D}\mathbf{q}\,\mathcal{D}\mathbf{r}\;\exp \left\{\frac{i}{\hbar} S_{CG}\left[\mathbf{q},\mathbf{r}\right]\right\}\,,
\end{equation}
where the coarse-grained action  $ S_{CG}$ reads
\begin{align*}
    S_{CG}[\mathbf{q} ,\mathbf{r}]=\int_{-\infty}^{\infty}dt\;  r^i(t)&\biggl\{-m\,\ddot{q}^i(t)-\nabla^i V (\mathbf{q}(t))+e^2\nabla^i G[\mathbf{q}(t),\mathbf{q}(t)] \biggr.\\
& -e^2\left( \delta^{il} \frac{d}{d t}-\dot{q}^l(t)\nabla^i\right) \int_{-\infty}^{\infty}dt' \; G_{R}^{lj} \left[{\bf q}(t), {\bf q}(t'); t-t'\right] \, \dot{q}^j (t') \\
\biggl.+\;i \,\frac{e^2}{2}\int_{-\infty}^{\infty}dt' \; r^j(t')&\left( \delta^{ i l} \frac{d}{dt}-\dot{q}^l (t)\nabla^i\right)\left(\delta^{jm}\frac{d}{d t'} -\dot{q}^m (t')\nabla'^j\right) G_{H}^{lm} \left[ {\bf q}(t),{\bf q}(t'); t-t'\right]\biggr\}+\mathcal{O}(r^{3})
\end{align*}
after being expanded with respected to $\mathbf{r}$. The prime over
$\nabla^{j}$ denotes the partial differentiation over $q^{j}(t')$.
The retarded Green's function and the Hadamard function  are defined
respectively by
\begin{eqnarray}
    \hbar\,G_{R}^{ij}(x-x')&=&i\,\theta(t-t')\,\bigl<\left[A_{\mathrm{T}}^i(x),A_{\mathrm{T}}^j(x')\right]\bigr>\,,\label{commutator}\\
    \hbar\,G_{H}^{ij}(x-x')&=&\frac{1}{2}\,\bigl<\left\{A_{\mathrm{T}}^i(x),A_{\mathrm{T}}^j(x')\right\}\bigr>\,.\label{anticommutator}
\end{eqnarray}
Next, we introduce the auxiliary noise fields $\xi^i (t)$ with the
Gaussian distribution,
\begin{equation}
    \mathcal{P}[\xi^i(t)]=\exp\left\{-\frac{\hbar}{2}\int_{-\infty}^{\infty}dt\int_{-\infty}^{\infty}dt'\;\left[\xi^i(t)\,G_{H}^{ij}{}^{-1}\left[{\bf q}(t),{\bf q}(t');t-t'\right]\,\xi^j(t')\right]\right\}\,,
\label{noisedistri}
\end{equation}
and thus the imaginary part of the coarse-grained action can be
expressed as a functional integration over $\xi_i (t)$ weighted by
the distribution function $ \mathcal{P}[\xi_i (t)]$. As a result, we
end up with
\begin{align*}
    &\quad\exp\left\{\frac{i}{\hbar}S_{CG}[\mathbf{q},\mathbf{r}]\right\}\\
&=\int\mathcal{D}
\xi_i \; \mathcal{P} [\xi_i (t)] \exp\left\{\frac{i}{\hbar} \left[ \mathrm{Re}\{S_{CG}\left[ {\bf q} , {\bf r} \right] \}-\hbar\,e \int_{-\infty}^{\infty}dt\; r^i \left( \delta^{ij}\frac{d}{d t} - q^j(t)\nabla^i \right)\xi^j \right]\right\} \,. \label{effectaction}
\end{align*}
The expressions in the squared brackets on the right hand side is
defined as the stochastic effective action, which consists of the
real part of the coarse-grained effective action as well as the
coupling term of the relative coordinate $r^i$ with the stochastic
noise $\xi^i$.

The Langevin equation is obtained by extremizing the
stochastic effective action and then setting $r^{i}$ to zero. By
doing so, we have ignored intrinsic quantum fluctuations of the
particle, and that holds as long as the resolution of the
measurement on length scales is greater than its position
uncertainty. The Langevin equation is then given by
\begin{eqnarray}
m \ddot{q}^i &+& \nabla^i V(\mathbf{q}(t)) + e^2\nabla^iG [{\bf q}(t),{\bf q}(t)]+e^2\left(\delta^{i l} \frac{d}{dt} - \dot{q}^l (t) \nabla_i\right) \nonumber \\
&\times &\int_{-\infty}^{\infty}dt' \; G_{R}^{lj} \left[{\bf q}(t),{\bf q}(t'); t-t'\right]\,\dot{q}^j (t') = -\hbar\, e \, \left( \delta^{il} \frac{d}{d t} - \dot{q}^l (t)\nabla^i \right) \, \xi^l (t)\label{nonlinearlangevin}
\end{eqnarray}
with the noise-noise correlation functions,
\begin{equation}
\langle\xi^i(t)\rangle =0\,,\qquad\qquad\langle\xi^i (t)\xi^j(t')\rangle=\frac{1}{\hbar} G_{H}^{ij} \left[{\bf q}(t), {\bf q}(t');
t-t'\right]\, . \label{noisecorrel}
\end{equation}
This Langevin equation encompasses fluctuation and
dissipation effects on the charge's motion from quantized
electromagnetic fields via the kernels $G_{H}^{ij}$ and $G_{R}^{ij}$
respectively, both of which are in turn linked by the
fluctuation-dissipation relation~\cite{KU}. The
fluctuation-dissipation relation is known to play a pivotal role in balancing
these two effects in order to dynamically stabilize the
nonequilibrium evolution of the particle under a fluctuating
environment. Mathematically, it relates the Fourier transform of the
fluctuation kernel $G_{H}^{ij}$ to the imaginary part of the
retarded kernel $G_{R}^{ij}$ as follows
\begin{equation}
    G_{H}^{ij}[{\bf q}(t),{\bf q}(t');\omega]=\operatorname{Im}\left\{G_{R}^{ij}[{\bf q}(t),{\bf q}(t');\omega]\right\}\coth\left[\frac{\beta\omega}{2}\right]\,. \label{df-T}
\end{equation}
In the zero temperature limit, the relation reduces to
\begin{equation}
    G_{H}^{ij}[{\bf q}(t),{\bf q}(t');\omega]=\operatorname{Im}\left\{G_{R}^{ij}[{\bf q}(t),{\bf q}(t');\omega]\right\}\left[\theta(\omega)-\theta(-\omega)\right]\,.\label{df-vac}
\end{equation}

It is found that the backreaction kernel functions of electromagnetic
fields in the Langevin equation \eqref{nonlinearlangevin} appear purely classical due to the fact that the coupling between the charge and electromagnetic potentials is
linear. The noise-noise correlation functions can in principle be
computed by taking an appropriate statistical average with the
distribution functional $\mathcal{P}[\xi^{i}(t)]$.
It is also seen from Eq.~\eqref{nonlinearlangevin} that the influence of electromagnetic fields takes the form of an
integral of the dissipation kernel over the past history of the
charge's trajectory, as well as a stochastic noise $\boldsymbol{\xi}$,
which drives the charge into a fluctuating motion. As it stands,
this is a nonlinear Langevin equation with non-Markovian
backreaction, and the noise depends in a complicated way on the
charge's trajectory because the noise correlation function itself is
a functional of the trajectory.

The general solution $q^{i}$ of the  Langevin equation can
be expressed as its mean value $q^{i}_{h}$ and a small deviation
$\delta q^{i}$ from the mean. Expanding the equation with respect to
$\delta q^i$ and then keeping its linear terms, we
may decompose the stochastic equation into the equations of motion
for $q^{i}_{h}$ and $\delta q^i$, respectively. The mean
trajectory $q^{i}_{h}$ satisfies the homogeneous part of the
Langevin equation, which describes the purely classical effects. On
the other hand, the equation  for the position fluctuations $\delta
q^{i}$ involves the stochastic noise $\xi^i$. The noise-driven
position fluctuations $\delta q^{i}$  thus are entirely of quantum
origin as seen from  an explicit $\hbar$ dependence
in the noise term. The backreaction dissipation effect on the evolution of $\delta q^{i}$
is expected to balance with the effect from the accompanying
stochastic noise via a fluctuation-dissipation relation where both effects
are of quantum nature~\cite{HU1}. This issue will be further studied
below.

\section{Stochastic Lorentz forces with the boundary}\label{sec2}
The integro-differential equation~\eqref{nonlinearlangevin} can be
cast into a form similar to the Lorentz equation. We consider a
charged particle moving in the vicinity of a perfectly conducting
plate. Let the plate be located at the $z=0$ plane. Then, the
tangential components of the electric field $\mathbf{E}$ and the
normal component of the magnetic field $\mathbf{B}$ on the plate
surface should vanish such that the boundary conditions of the vector
potential $\mathbf{A}$ are given by
\begin{equation}
    A_0=0\,,\qquad\text{and}\qquad A_x=A_y=0\,,
\end{equation}
leading to
\begin{equation}
    \frac{\partial{A}_z}{\partial z}=0
\end{equation}
as the result of the Coulomb gauge. The transverse vector potential
$\mathbf{A}_{\mathrm{T}}$ in the $z>0$ region is given by,
\begin{eqnarray}
    \mathbf{A}_{\mathrm{T}}(x)&=&\int\!\frac{d^2\mathbf{k}_{\parallel}}{2\pi}\!\int_0^{\infty}\!\frac{dk_z}{(2\pi)^{1/2}}\;\frac{2}{\sqrt{2\omega}}\biggl\{a_1(\mathbf{k})\,\hat{\mathbf{k}}_{\parallel}\times\hat{\mathbf{z}}\,\sin k_zz\biggr.\nonumber \\
 &&\quad\quad\quad\biggl.+\;a_2(\mathbf{k})\left[i\,\hat{\mathbf{k}}_{\parallel}\left(\frac{k_z}{\omega}\right)\sin k_zz -\hat{\mathbf{z}}\left(\frac{k_{\parallel}}{\omega}\right)\cos k_zz\right]\biggr\}\;e^{i\mathbf{k}_{\parallel}\cdot\mathbf{x}_{\parallel}-i\omega t}+\text{H.C.}\,,
\end{eqnarray}
where the circumflex identifies unit vectors. The position vector
$\mathbf{x}$ is the shorthanded notation of
$\mathbf{x}=(\mathbf{x}_{\parallel},z)$, where
$\mathbf{x}_{\parallel}$ is the components parallel to the plate.
Similarly, the wave vector is expressed by  $\mathbf{k}=(
\mathbf{k}_{\parallel},k_ z)$ with $\omega^2=k_{\parallel}^2+k_z^2$.
The commutation relations of the creation and annihilation operators
are satisfied by
\begin{equation}
    [a_{{\lambda}^{\vphantom{l}}}^{\vphantom{\dagger}}
    (\mathbf{k}),a_{\lambda'}^{\dagger}(\mathbf{k}')]=\delta_{{\lambda}{\lambda}'}\,
    \delta(\mathbf{k}_{\parallel}^{\vphantom{'}}-\mathbf{k}'_{\parallel})\,
       \delta(k_z^{\vphantom{'}}-k'_z)\,,\qquad\text{with $\lambda,\lambda'=1,2$}\,,
\end{equation}
and are zero otherwise. Then, the retarded Green's function and
Hadamard function can be explicitly expressed as the sum of the
free-space and the boundary-induced contributions,
\begin{eqnarray}
G_{R}^{ij} (\mathbf{q}(t),\mathbf{q}(t');t-t') &=& i\, \theta (t-t') \int \frac{d^3 \mathbf{k}}{(2\pi)^3}\;\frac{1}{2\omega}\left\{\left(\delta^{ij}-\frac{k^{i}k^{j}}{\omega^2}\right)e^{i\mathbf{k}\cdot[\mathbf{q}(t)-\mathbf{q}(t')]}\right.\label{ret} \\
  &-&\left.\left(\delta^{ij}-\frac{k^{i}k^{j}}{\omega^2}-2\hat{z}^{i}\hat{z}^{j}+2\frac{k_z}{\omega}\hat{k}^{i}\hat{z}^{j}\right)e^{i\mathbf{k}\cdot[\mathbf{q}(t)-\overline{\mathbf{q}}(t')]}\right\}\left( e^{-i\omega(t-t')}-\rm{C.C.} \right)\,,\nonumber \\
G_{H}^{ij}(\mathbf{q}(t),\mathbf{q}(t');t-t') &=& \frac{1}{2}\int \frac{d^3 \mathbf{k}}{(2 \pi )^3}\;\frac{1}{2\omega}\,\coth \left[\frac{\beta k}{2}\right]\left\{\left(\delta^{ij}-\frac{k^{i}k^{j}}{\omega^2}\right)e^{i \mathbf{k}\cdot[\mathbf{q}(t)-\mathbf{q}(t')]}\right.\label{Hadamrd} \\
  &-&\left.\left(\delta^{ij}-\frac{k^{i}k^{j}}{\omega^2}-2\hat{z}^{i}\hat{z}^{j}+2\frac{k_z}{\omega}\hat{k}^{i}\hat{z}^{j}\right)e^{i\mathbf{k}\cdot[\mathbf{q}(t)-\overline{\mathbf{q}}(t')]}\right\}\left(e^{-i\omega(t-t')}+\rm{C.C.} \right) \,,  \nonumber
\end{eqnarray}
where the position $\overline{\mathbf{q}}$ denotes the location of
the mirror image of the original charge at $\mathbf{q}$ with respect
to the boundary at the $z=0$ plane, and they are related by $\overline{q}^{i}=(\delta^{ij}-2
\hat{z}^{i}\hat{z}^{j})\,q^{j}$.

The integral expression in Eq.~\eqref{nonlinearlangevin} can be
realized in terms of   the Li\'enard-Wiechert potential in the
Coulomb gauge due to a moving charge,
\begin{equation}
A^{i}_{\rm LW}(\mathbf{q})=\int d^4
x'\;G^{ij}_{R}[\mathbf{q}(t),x']\,j^{j}(x';{\bf q}(t'))\,.
\end{equation}
This potential $\mathbf{A}_{\rm LW} (\mathbf{q})$ clearly depends on
the past history of the charge's motion~\cite{JD}.  Together with
the instantaneous Coulomb potential term, the backreaction can be
expressed in a gauge invariant way to necessarily maintain
underlying gauge symmetry, respected by the Lagrange we begin with.
With the definition of the current density in
Eq.~\eqref{charge-current}, the straightforward algebraic
manipulation further shows that the Langevin equation can be
re-expressed as retarded Lorentz forces and the stochastic
components,
\begin{equation}
m \ddot{q}^{i}+\nabla_{\mathbf{q}}^{i} V(\mathbf{q}) = e
\left[E^i(\mathbf{q})+\epsilon_{ijk}\,\dot{q}^j(t)B^k
(\mathbf{q})\right]-\hbar\, e \left(\delta^{ij} \frac{d}{d t} -
\dot{q}^j(t)\nabla_{\mathbf{q}}^j\right)\xi^j(t)\,.\label{gaugeinvaraintLangevin}
\end{equation}
The electromagnetic fields are defined by
\begin{equation}
\mathbf{E}= -\frac{\partial}{\partial t}\mathbf{A}_{\rm
LW}-\nabla_{\mathbf{q}}G\,,\qquad\qquad\qquad\mathbf{B}=\nabla_{\mathbf{q}}\times\mathbf{A}_{\rm
LW}\,.
\end{equation}
Next we write electromagnetic fields in reference to the retarded
spacetime coordinates and explicitly have them divided into the
free-space contribution and the correction out of the boundary.
Let the free-space part $G_{R}^{ij;(0)}$ and the boundary
correction $G_{R}^{ij;(b)}$ of the retarded Green's function be
respectively given by
\begin{align*}
     G_{R}^{ij;(0)}(\mathbf{q},\mathbf{q}';\tau)&=i\,\theta(t-t')\int\frac{d^3\mathbf{k}}{(2\pi)^3}\;\frac{1}{2\omega}\left\{\delta^{ij}-\frac{k^{i}k^{j}}{\omega^2}\right\}e^{i\mathbf{k}\cdot(\mathbf{q}-\mathbf{q}')-i\omega\tau}+\rm{C.C.}\,,\\
     G_{R}^{ij;(b)}(\mathbf{q},\mathbf{q}';\tau)&=-i\,\theta(t-t')\int\frac{d^3\mathbf{k}}{(2\pi)^3}\;\frac{1}{2\omega}\left\{\delta^{ij}-\frac{k^{i}k^{j}}{\omega^2}-2\hat{z}^{i}\hat{z}^{j}+2\frac{k_z}{\omega}\hat{k}^{i}\hat{z}^{j}\right\}e^{i\mathbf{k}\cdot(\mathbf{q}-\overline{\mathbf{q}}')-i\omega\tau}+\rm{C.C.}\,,
\end{align*}
where $\mathbf{q}'=\mathbf{q}(t')$ and $\tau=t-t'$. The similar
decomposition is applied to the instantaneous Coulomb potential.
Therefore, we end up with
\begin{equation}
\mathbf{E}=\mathbf{E}^{(0)}(\mathbf{q},\mathbf{R})+\mathbf{E}^{(b)}(\mathbf{q},\mathbf{R})\,,\qquad\qquad\mathbf{B}=\left[\mathbf{n}\times\mathbf{E}^{(0)}(\mathbf{q},\mathbf{R})+\overline{\mathbf{n}}\times\mathbf{E}^{(b)}(\mathbf{q},\mathbf{R})\right]_{\mathrm{ret}}\,,
\label{retarded-lorentz-f}
\end{equation}
where
\begin{align}
     \mathbf{E}^{(0)}(\mathbf{q},\mathbf{R})&=e\,\left[\frac{\mathbf{n}-\dot{\mathbf{q}}}{\gamma^{2}(1-\dot{\mathbf{q}}\cdot{\mathbf{n}})^{3}\,{R}^{2}}\right]_{\rm{ret}}+e\,\left[\frac{\mathbf{n}\times[(\mathbf{n}-\dot{\mathbf{q}})\times\ddot{\mathbf{q}}]}{(1-\dot{\mathbf{q}}\cdot\mathbf{n})^{3}\,R}\right]_{\rm{ret}}\,,\label{E}\\
     \mathbf{E}^{(b)}(\mathbf{q},\mathbf{R})&=-\,e\,\left[\frac{\overline{\mathbf{n}}-\dot{\overline{\mathbf{q}}}}{\overline{\gamma}^{2}(1-\dot{\overline{\mathbf{q}}}\cdot{\overline{\mathbf{n}}})^{3}\,\overline{R}^{2}}\right]_{\rm{ret}}-e\,\left[\frac{\overline{\mathbf{n}}\times[(\overline{\mathbf{n}}-\dot{\overline{\mathbf{q}}})\times\ddot{\overline{\mathbf{q}}}]}{(1-\dot{\overline{\mathbf{q}}}\cdot\overline{\mathbf{n}})^{3}\,\overline{R}}\right]_{\rm{ret}}\,,\label{retarded-E}
\end{align}
in which $\mathbf{R} =\mathbf{q}(t)-\mathbf{q}(t_{R})$,
$\mathbf{n}=\mathbf{R}/R$, and
$\gamma=(1-\dot{{\mathbf{q}}}^{\,2})^{-1/2}$. The retarded time
$t_{R}$ is defined as $ t_{R}= t-R$. The variables with the overbar
are given by replacing the source point $\mathbf{q}(t_{R})$ with its
image position $\overline{\mathbf{q}}(t_{R})$ in their definition.
The minus sign in the definition of the retarded electric
field~\eqref{retarded-E} implies that this field can be interpreted
as  radiation out of the image charge due to the boundary. Thus
Eqs.~\eqref{retarded-lorentz-f}-\eqref{retarded-E} are consistent
with what would be obtained for electromagnetic fields due to a
moving charge and its image~\cite{JD}.

The free-space part of the Li\'enard-Wiechert potentials involves
its associated retarded Green's function, which is nonvanishing only for
lightlike spacetime intervals. Since the worldline of a massive
particle is timelike,  in the absence of the boundary the charge at
the present time can not be affected by backreaction from previously
emitted radiation due to the nonuniform motion of the charge
itself~\cite{JD}; however, radiation can  backreact on the charge at
the time when it is just emitted. Thus, the nonlocal form of the free-space contribution of the Li\'enard-Wiechert potential
reduces to a purely local effect. Besides it may  lead to
short-distance divergence in the coincidence limit. This ultraviolet
divergence arises from the assumption that the point-like particle
interacts with fields, and must be regularized to have a finite and
unambiguous result. Then the divergent part is absorbed by mass
renormalization, while the finite backreaction takes the form of electromagnetic self-forces, given by a third-order time derivative of the position. It may bring about issues such as the runaway solutions and acausality on the dynamical
evolution of a point charge~\cite{RO}.

Needless to say, the emitted radiation may backscatter off the
boundary, and in turn affects the charge's motion at a later time.
Thus, this backreaction owing to the presence of the boundary
depends on the past history of the charge's trajectory,  leading to
the memory effect. This non-Markovian processes can also be
understood as if radiation was emitted at a retarded time from the
image charge at $\overline{\mathbf{q}}$, which is the location of
the mirror reflection of the point charge at $\mathbf{q}$ due to the
presence of the conducting plate. The lag corresponds to
the time delay for radiation to travel from the image counterpart at
an earlier time to the charge itself at the present time, roughly
being equal to the round-trip traveling time of radiation between the plane boundary and the
charge.

In addition, from the interpretation of the stochastic noise and the noise-noise correlation, we may formally identify $-\hbar\,\xi^{i}$
as the stochastic vector potentials $A^i_s$ such that their correlation functions are
\begin{equation}
    \langle A^i_s(t)\rangle=0\,,\qquad\qquad \langle A^i_s(t)A^j_s(t')\rangle=\hbar\,G^{ij}_H[\mathbf{q}(t),\mathbf{q}(t');t-t']\,.
\end{equation}
The noise term on the right hand side of
Eq~\eqref{gaugeinvaraintLangevin} can then be thought of as the
stochastic Lorentz force $F_s$,
\begin{equation}
    F^i_s=e\left[E^i_s+\epsilon_{ijk}\,\dot{q}^j(t)B_s^k\right]\,,
\end{equation}
where
\begin{equation}
    E^i_s=-\frac{\partial}{\partial t}A^i_s\,,\qquad\text{and}\qquad B^i_s=-\nabla_{\mathbf{q}}^i\times A^i_s\,.
\end{equation}
The origin of the stochastic Lorentz force by construction comes
from  electromagnetic vacuum fluctuations. Here we note that the
stochastic electromagnetic fields $(E^i_s,B^i_s)$ involve only the
transverse components of gauge potentials. In the Coulomb gauge, the instantaneous Coulomb potential, which is determined by the Gauss law, is not a dynamical variable when the intrinsic quantum fluctuations of the charge are ignored, and hence it has no corresponding stochastic component. In the end, the
Langevin equation~\eqref{gaugeinvaraintLangevin} for the point charge interacting with quantized electromagnetic fields in the presence of the conducting plate can
be nicely cast into a gauge-invariant form,
\begin{align}
    m \ddot{q}^{i}+\nabla_{\mathbf{q}}^{i}V(\mathbf{q})
    = e\left[E^{i}(\mathbf{q})+\epsilon_{ijk}\,\dot{q}^j(t)B^{k}(\mathbf{q})\right]+e\left[E^i_s+\epsilon_{ijk}\,\dot{q}^j(t)B_s^k\right]\, . \label{fullLorentz}
\end{align}
 Then the expressions in the first pair of squared brackets on the
right hand side denote the retarded electromagnetic forces, while
the terms in the second pair are its stochastic components,
manifested from the quantum fluctuations of electromagnetic fields.

\section{Langevin equation under dipole approximation and its solution}\label{sec4}
The nonlinear, non-Markovian Langevin equation is far too
complicated to study further without any approximation. We will
consider that a charged particle undergoes the harmonic
motion, and assume that the amplitude of oscillation is sufficiently
small. The appropriate approximation for a non-relativistic motion
will be the dipole approximation. This approximation amounts to
considering the backreaction solely from  electric fields, and
linearizing the Langevin equation in such a way that the equation of
motion remains non-Markovian. Since the presence of the boundary
will anisotropically modify the effects of electromagnetic fields on
the evolution of the charge, we will consider the motion in two
different directions, that is, either parallel or perpendicular to
the plane boundary.

\subsection{parallel motion}
When the charged particle moves parallel to the boundary, say, in
the $x$ direction, let the equilibrium point be located at the
coordinates $(x,y,z)=(0,0,z_0)$. The linearized Langevin equation
for the motion displaced from its equilibrium position reduces to
\begin{equation}\label{xxx}
    m\ddot{q}^x(t)+\partial_{x}^{2}V(z_0)\,q^x(t)+e^2\int_0^tdt'\;\dot{g}_{R}^{x}[z_0,z_0;t-t']\,\dot{q}^x(t')=-\hbar\,e \,\dot{\xi}^x(t)\,,
\end{equation}
from Eqs.~\eqref{nonlinearlangevin}, \eqref{ret}, and \eqref{Hadamrd} under the
dipole approximation. Here $q^{i}$ denotes the displacement from the equilibrium point. The $xx$ component of the retarded Green's
function in the dipole approximation is denoted by $g_{R}^{x}[z_0,z_0;t-t']$, and can
be written in terms of the spectral density $\rho^{x}$ as
\begin{equation}
    g_{R}^{x}[z_0,z_0;t-t']=-\theta(t-t')\int_0^{\infty}\frac{dk}{\pi}\;\rho^{x}(z_0,z_0;k)\sin[k(t-t')]\,,
\end{equation}
where the spectral density is given by
\begin{equation}
    \rho^{x}(z_0,z_0;k)=-\frac{k}{\pi}\left[\frac{1}{3}-\frac{\sin(2k
z_0)}{2(2k z_0)}-\frac{\cos(2k z_0)}{2(2k
z_0)^2}+\frac{\sin(2k z_0)}{2(2k z_0)^3}\right]\, .
\end{equation}
The instantaneous Coulomb potential in this case is found to have no
effect on the motion in the $x$ direction, but will establish a
static attraction force between the charge and its image in the $z$
direction. Hence, the applied potential $V$ is assumed to have an
additional component, other than the harmonic potential, to counteract the static force so that the motion of the charge remains on the $z=z_{0}$ plane. In addition, the
motion has been assumed to start at $t=0$, and its initial
conditions are chosen to be $q^{x}(t<0)=q^{x}(0)$ and
$\dot{q}^{x}(t<0)=0$. In general, it is insufficient for a
non-Markovian equation to have a unique solution if a finite number
of initial conditions are specified. Thus in this case, the initial
conditions are given over half of the real axis $t<0$, although later
it will become clear that the requirement on specifying initial
conditions can be less stringent in the current case. Physically,
these initial conditions can be achieved by applying an appropriate
external constraint to hold the particle at rest at the position
$q^{x}(0)$ for $t<0$. Then the applied potential is suddenly
switched off to $ V(\mathbf{q})$ at the time $t=0$ so that the
dynamics of the particle at later times follows the Langevin
equation. The accompanying noise-noise correlation functions due to
quantum fluctuations of $E^{x}$ fields are given by
\begin{equation}
    \langle\xi^{x}(t)\rangle=0\,,\qquad\qquad\langle\xi^{x}(t)\xi^{x}(t')\rangle=\frac{1}{\hbar}\,g_{H}^{x}[z_0,z_0;t-t']\,,
\end{equation}
where
\begin{equation}
    g_{H}^{x}[z_0,z_0;t-t']=-\int_0^{\infty}\frac{dk}{2\pi}\;\rho^{x}(z_0,z_0;k)\,\cos[k(t-t')] \,. \label{hadama-x}
\end{equation}
The noise kernel $g_{H}^{x}$ is related to the dissipation kernel
$g_{R}^{x}$ via a fluctuation-dissipation relation under the dipole
approximation.

Carrying out the integration over $k$ makes the Langevin
equation physically more transparent,
\begin{align}
     m_{r}\ddot{q}^{x}(t)+\left[m_{r}\omega_{0}^{2}+\frac{e^{2}}{4\pi}\frac{1}{(2z_{0})^3}\right]& q^{x}(t)=-\hbar\,e \,\dot{\xi}^x(t)+\frac{e^2}{6\pi}\dddot{q}^{x}(t)\label{time-delay-xx}\\
     &+\frac{e^{2}}{4\pi}\left[\frac{\ddot{q}^{x}(t-T)}{(2z_0)}+\frac{\dot{q}^{x}(t-T)}{(2z_0)^{2}}+\frac{q^{x}(t-T)}{(2z_0)^{3}}\right]\,,\notag
\end{align}
with the time delay $ T=2z_{0}/c $ and the renormalized mass $
m_{r}=m+ e^2 \Lambda /3\pi^2 .$ Here for $t>0$ the charged particle
is set into harmonic oscillation with frequency $\omega_0$ by
specifying the proper potential $V(\mathbf{q})$. The ultraviolet
divergence arises due to summing up backreaction effects
from the free-space contribution over all energy scales of fields in
the coincidence limit. This type of divergence is often seen in
quantum theory as well as classical theory when an infinite number
of degrees of freedom are involved. The energy cutoff $\Lambda$ is
then introduced to regularize the integral. The cutoff scale can in
principle be chosen to be the inverse of the particle's classical
radius. However, this energy scale is already well beyond the regime
of the validity of a nonrelativistic, single-particle description for the Langevin
equation. Thus, instead, the more sensible energy cutoff scale will
be the inverse of the width of the charge's wavefunction. It
essentially quantifies the intrinsic uncertainty on the charged
particle. The divergence is then absorbed into mass renormalization
as shown above. The finite backreaction effect is given by the
third-order time derivative of the position.

The boundary-induced backreaction in the squared brackets on the
right hand side of Eq.~\eqref{time-delay-xx} reveals the
non-Markovian nature in the argument of each individual term. The
assumption of small amplitude for the charged oscillator simplifies
a general non-Markovian integro-differential equation into an
ordinary differential equation with a fixed time delay. This is
expected due to the fact that by small amplitude, we mean the
variation of the position for the charge oscillator is much smaller
than its distance to the plate $z_{0}$. It implies that the time
variation owing to the oscillation amplitude is ignorable compared
to the time $2z_{0}/c$ needed for the radiation to have a round trip
between the charge and the boundary. Therefore, the dominant
contribution of the time delay will be just given by a fixed
constant $T=2z_{0}/c$. The term with $\ddot{q}^{x}(t-T)$ is derived
from the acceleration field of  electric fields while the rest two
terms come from the velocity field. The time difference $t-T$
indicates that the backreaction effect due to the boundary on the
charge at the time $t$ depends on the charge's dynamics at an
earlier time $t-T$. The time delay $2z_0/c$ may also be thought of
as the traveling time for radiation emitted from the image charge at
an earlier time to reach the charge at a later time. The memory
effect is thus described by such a time delay differential equation.
This equation can also be obtained from
Eq.~\eqref{gaugeinvaraintLangevin} by expanding the expression of
the retarded Lorentz force to  the linear terms in $q^{x}$,
$\dot{q}^{x}$ and $\ddot{q}^{x}$ in the non-relativistic limit.

In order to find the solutions, it apparently requires the knowledge of $q^x$ at
earlier times. Since the Langevin equation under consideration
reduces to an ordinary differential equation with a fixed time delay
$T$, the specification of the initial conditions can be relaxed to the conditions
\begin{equation}\label{initialcon}
    q^{x}(-T<t<0)=q^{x}(0)\,,\qquad\qquad\qquad\dot{q}^{x}(-T<t<0)=0\,,
\end{equation}
in the interval $-T<t<0$, instead of the whole negative real axis of $t$.
Then the equation can be solved by means of the Laplace
transform. From Eq.~\eqref{xxx}, the Laplace transformed
Langevin equation reads
\begin{equation}
m_{r}\left[s^2\widetilde{q}^{x}(s)-s\,q^{x}(0)+\omega^2_0\widetilde{q}^{x}(s)\right]+e^2\widetilde{g}'^{x}_{R}[z_0,z_0;s]\left[s^2\widetilde{q}^{x}(s)-s\,q^{x}(0)\right]=-
\hbar e \widetilde{\xi}^{x}_{s}(s)\,,
\end{equation}
where the initial conditions~\eqref{initialcon} have been applied.
The Laplace transformation of a function $f(t)$ is defined by
\begin{equation*}
    \widetilde{f}(s)=\int_0^{\infty}dt\;e^{-st}f(t)\,,\qquad\qquad t>0\,,
\end{equation*}
while the inverse Laplace transform of $\widetilde{f}(s)$ is
\begin{equation*}
    f(t)=\frac{1}{2\pi
    i}\int_{\mathcal{C}}ds\;\widetilde{f}(s)\,e^{st}\, ,
\end{equation*}
where the Bromwich contour $\mathcal{C}$ counterclockwisely
encompass all singularities of $\widetilde{f}(s)$ on the complex $s$
plane. Since ultraviolet divergence has been absorbed into mass renormalization, in order to avoid being cluttered with the notations all variables will be assumed renormalized from now on unless specified otherwise. The renormalized kernel $\widetilde{g}^{x}_{R}[z_0,z_0;s]$ is given by
\begin{equation}
\widetilde{g}^{x}_{R}[z_0,z_0;s]=-\int^{\Lambda} \frac{dk}{\pi}\;\rho^{x} (z_0,z_0;k)\,\frac{k}{k^{2}+s^{2}}-\frac{\Lambda}{3\pi^2}\,, \label{re-retarded-g}
\end{equation}
The solution to the Laplace transformed equation is readily obtained as
\begin{equation}
\widetilde{q}^{x}(s)=\frac{s\left\{1+\frac{e^2}{m}\,\widetilde{g}_{R}^{x}[z_0,z_0;s]
\right\}q^{x}(0)-\frac{\hbar \,
e}{m}\,\widetilde{\xi}^{x}_{s}(s)}{s^2+\omega_0^2
+\widetilde{\Sigma}_{x}(s)}\,,
\end{equation}
where the $\widetilde{\Sigma}_{x}(s)$ kernel is
\begin{equation}
    \widetilde{\Sigma}_{x}(s)=\frac{e^2}{m}\,s^2\,\widetilde{g}_{R}^{x}[z_0,z_0;s]\,.\label{sigma-def}
\end{equation}
Thus, the solution in the time domain can be given by the inverse
Laplace transformation of $\widetilde{q}^{x}(s)$, and it reads in
terms of the mean trajectory and its deviation from the mean,
\begin{equation}
     q^{x}(t)=\left[\dot{K}_{x}(t)+\frac{e^2}{m}\int_0^tdt'\;g_{R}^{x}[z_0,z_0;t-t']\,\dot{K}_{x}(t')\right]q^{x}(0)-\frac{e\hbar}{m}\int^t_0dt'\;K_x(t-t')\,\dot{\xi}^{x}(t')\,,\label{realtimesol}
\end{equation}
where the kernel $K_x(t)$ is the inverse Laplace transform of $\widetilde{K}_{x}(s)=[s^2+\omega_0^2 +\widetilde{\Sigma}_{x}(s)]^{-1}$,
\begin{equation}\label{invLap}
K_x(t)=\int_{\mathcal{C}}ds\;\frac{1}{s^2+\omega_{0}^2+\widetilde{\Sigma}_{x}(s)}\,e^{st}\,,
\end{equation}
with $K_{x}(0)=0$ and $\dot{K}_{x}(0)=1$. It is seen from
Eqs.~\eqref{re-retarded-g} and \eqref{sigma-def} that the kernel $\widetilde{\Sigma}_{x}(s)$ has a
branch-cut along the imaginary $s$ axis, so the discontinuity of
the kernel $\widetilde{\Sigma}_{x}(s)$ over the branch-cut can be made
explicitly by letting $s=i\omega+0^{\pm}$. In terms of $\omega$, the
kernel $\widetilde{\Sigma}_{x}$ becomes
\begin{equation}
    \widetilde{\Sigma}_{x}(s=i\,\omega+0^{\pm})=\operatorname{Re}\Sigma_{x}(\omega)\pm i\operatorname{Im}\Sigma_{x}(\omega)\,,
\end{equation}
where
\begin{align}
    \operatorname{Re}\Sigma_{x}(\omega)&=\frac{e^2}{m}\,\omega^2\left[\int^{\Lambda}\frac{dk}{\pi}\;k\, \rho^{x}(z_0,z_0;k)\,\mathcal{P}\left(\frac{1}{k^2-\omega^2}\right)+\frac{\Lambda}{3\pi^2} \right] \, \\
     \operatorname{Im}\Sigma_{x}(\omega)&=-\frac{e^2}{m}\,\operatorname{sgn}(\omega)\,\omega^2\int^{\infty}_{0}dk\;k\,\rho^{x}(z_0,z_0;k)\,\delta(k^{2}-\omega^2)\,.
\end{align}
Carrying out the integrals yields
\begin{eqnarray}
    \operatorname{Re}\Sigma_{x}(\omega)&= &\frac{e^2}{4\pi m}\,\omega^3\left[\frac{1}{(2\omega z_{0})^{3}}+\frac{\cos(2\omega z_{0})}{(2\omega z_{0})}-\frac{\sin(2\omega z_{0})}{(2\omega z_{0})^{2}}-\frac{\cos(2\omega z_{0})}{(2\omega z_{0})^{3}}\right]\,, \label{resigma-x} \\
    \operatorname{Im}\Sigma_{x}(\omega)&= &\frac{e^2}{4\pi m}\,\operatorname{sgn}(\omega)\,\omega^3\left[\frac{2}{3}-\frac{\sin(2\omega z_{0})}{(2\omega z_{0})}-\frac{\cos(2\omega z_{0})}{(2\omega z_{0})^{2}}+\frac{\sin(2\omega z_{0})}{(2\omega z_{0})^{3}}\right]\,. \label{imsigma-x}
\end{eqnarray}
In general, the branch-cut of the $\widetilde{\Sigma}_{x}(s)$ kernel
lies within the intervals $(\omega_{\rm{th}},\Lambda)$
and $(-\Lambda,-\omega_{\rm{th}})$, where $\Lambda$ is the
energy cutoff and $\omega_{\rm{th}}$ is the threshold energy above
which the process of particle creation is possible~\cite{PE}.
However, for quantized electromagnetic fields, its threshold energy vanishes due to masslessness of the photon, i.e. $\omega_{\rm{th}}=0$. Thus the branch-cut will extend
along the imaginary $s$ axis from $-i\Lambda$ to $i\Lambda$.

The locations of the poles on $1/(s^2+\omega_0^2
+\widetilde{\Sigma}_{x}(s))$ can be found by solving the following
equation,
\begin{equation}\label{poleeq}
 s^2+\omega_0^2+\widetilde{\Sigma}_{x}(s)=0\,.
\end{equation}
This pole equation turns out to be a transcendental function of $s$
due to the presence of the boundary. It may have an infinite number
of solutions. This in turn implies that the integro-differential
equation can be viewed as an infinite order of the ordinary
differential equation. Thus an infinite number of initial conditions
are needed~\cite{BE}.
\begin{figure}
\centering
    \scalebox{0.9}{\includegraphics{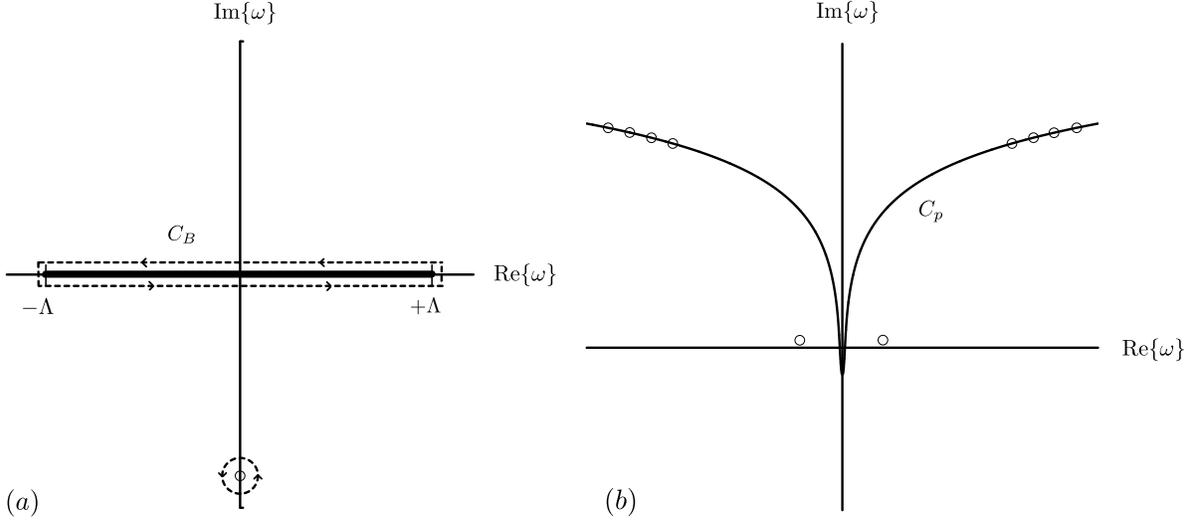}}
    \caption{(a) On the first Riemann sheet, the branch-cut lies along the real axis of $\omega$ and a runaway pole sits at the negative imaginary axis. All the other poles stay on the second Riemann sheet. Thus, the Bromwich contour reduces to two straight lines parallel but infinitesimally close to the cut from both sides if the contribution of the runaway pole is discarded. (b) The resonance modes are found asymptotically in the complex $\omega$ plane along the curve $ C_p$.}\label{Fi:branch}
\end{figure}
For the weak coupling constant $e^2$,  perturbatively, we may
let $s=i(\omega_0+\delta\omega)+\delta\gamma$, where
$\delta\omega-i\,\delta\gamma$ is the shift of the pole owing to interaction with electromagnetic fields. It is then plugged into the equation~\eqref{poleeq} to find
$\delta\omega$ and $\delta\gamma$. We assume $\delta\gamma>0$ for
the moment. To the order $e^2$, the perturbative solution is found
to be
$\delta\omega=\operatorname{Re}\Sigma_{x}(\omega_0)/(2\omega_0)$,
and
$\delta\gamma=-\operatorname{Im}\Sigma_{x}(\omega_0)/(2\omega_0)$. However, since from Eq.~\eqref{imsigma-x}, $\operatorname{Im}\Sigma_{x}(\omega)>0$ for all
positive values of $\omega$, it leads to
$\delta\gamma<0$, inconsistent with the assumption we made earlier.
A similar contradictory result will be found through the same
argument by assuming $\delta\gamma<0$. As a result, the poles do
not exist in the first Riemann sheet, and may appear in the second
sheet. This is the known fact for the resonance~\cite{AL}. In fact, there are still an infinite number of poles lying on the upper half of the second Riemann sheet. However, one exception is the runaway pole, which resides on the positive real axis of the complex $s$ plane. Since the kernel $\widetilde{\Sigma}_{x}(s)$ is real when evaluated at the runaway pole, this pole stays on the first Riemann sheet. It corresponds to either a runaway motion of the charge or the acausal evolution due to preacceleration. Since it is viewed as unphysical if the asymptotically bounded motion is considered~\cite{JD}, we then discard the contributions of the runaway-type poles to the inverse Laplace
transformation~\eqref{invLap}.  Accordingly, the contour on the first
Rienmann sheet can be deformed to be parallel and infinitesimally
close to the branch-cut as shown in Fig.~\ref{Fi:branch}. The
real-time solution of $K_{x}(t)$ can be written as
\begin{equation}\label{k-int}
    K_x(t)=\int^{\infty}_{\omega_{\rm{th}}=0}\frac{d\omega}{2\pi}\;\frac{4\operatorname{Im}\Sigma_{x}(\omega)}{\left[\omega^2-\omega_0^2-\operatorname{Re}\Sigma_{x}(\omega)\right]^2+\left[\operatorname{Im}\Sigma_{x}( \omega)\right]^{2}}\,\sin(\omega t)\,,\qquad t>0\,.
\end{equation}
Similarly, the time derivative of the $K_{x}$ kernel is given by
\begin{equation}\label{kd-int}
    \dot{K}_x(t)=\int^{\infty}_{\omega_{\rm{th}}=0}\frac{d\omega}{2\pi}\;\frac{4\omega\operatorname{Im}\Sigma_{x}(\omega)}{\left[\omega^2-\omega_0^2-\operatorname{Re}\Sigma_{x}(\omega)\right]^2+\left[\operatorname{Im}\Sigma_{x}( \omega)\right]^{2}}\,\cos(\omega t)\,,\qquad t>0\,.
\end{equation}
The integrand in a weak coupling limit shows a Breit-Wigner
feature. The width of the resonance is related to the
imaginary part of the kernel $\Sigma_{x}$, and its peak is located at the resonance frequency $\Omega_{x}$ to be
determined later. For the sufficiently late times, $t\gg T$, the time derivative of $K_{x}(t)$ can be approximated by taking into account only the contribution of the resonance mode, and is given by
\begin{equation}
    \dot{K}_x (t)\sim Z_x\,\cos(\Omega_{x}t+\alpha_x)\,e^{-\Gamma_x t}\,,\qquad t>0\,, \label{realtimeK}
\end{equation}
where the resonance frequency $\Omega_{x}$ and the decay constant
$\Gamma_x$ are given by
\begin{equation}
\Omega_{x}\sim\omega_0+\frac{\operatorname{Re}\Sigma_{x}(\omega_0)}{2\omega_0}\,,\quad\quad \Gamma_x\sim Z_x\frac{\operatorname{Im}\Sigma_{x}(\Omega_x)}{2\Omega_x}\,, \label{frequency-decay}
\end{equation}
respectively. In addition, the phase shift $\alpha_x$ and $Z_x$ are
\begin{equation}
     Z_x\sim\left[1-\frac{\partial\operatorname{Re}\Sigma_{x}(\Omega_x)}{\partial\Omega_x^2}\right]^{-1}\,,\quad\quad\alpha_x\sim Z_x\frac{\partial\operatorname{Im}\Sigma_{x}(\Omega_x)}{\partial\Omega_x^2}\,.
\label{rewavefun-phase}
\end{equation}
The long-time dynamics of the kernel $\dot{K}(t)$ can also be
investigated by examining the low-frequency behavior of the
integrand in Eq.~\eqref{kd-int} along the branch-cut in the neighborhood of threshold
energy. When the resonance
peak is far away from threshold energy, $\omega_{\rm{th}}\neq\Omega$,
the imaginary part of the kernel $\Sigma$ generally behaves like
$\operatorname{Im}\Sigma\propto(\omega-\omega_{\rm{th}})^{n}$ as
$\omega$ approaches  $\omega_{\rm{th}}$. For nonzero threshold
energy, $\omega_{\rm{th}}\neq 0$, the integration over $\omega$ in
Eq.~\eqref{kd-int} leads to
\begin{equation}\label{massive}
    \dot{K}(t)\propto\int_{\omega_{\mathrm{th}}}^{\infty}d\omega\;(\omega-\omega_{\rm{th}})^{n}\cos(\omega t)=\frac{1}{t^{n+1}}\cos(\omega_{\rm{th}}t+\frac{n+1}{2}\pi)\,,
\end{equation}
which results in a power-law decay for its late-time
dynamics~\cite{DB}. However, for electromagnetic fields with  vanishing threshold energy, infrared photons
can be generated even with an infinitesimally small amount of
energy. The backreaction is expected to damp out the motion of the
 particle more effectively than that of the massive field.
Thus as shown in Eq.~\eqref{imsigma-x},  because $\operatorname{Im}\Sigma$ is proportional to
$\omega^{3}$ for small $\omega$, we find
\begin{equation}
    \dot{K}(t)\propto\int_{0}^{\infty}d\omega\;\omega^{4}\cos(\omega t)=0\,,
\end{equation}
after the proper regularization. This result can also be obtained
from Eq.~\eqref{massive}. The conclusion holds true for all
higher-order time derivatives of $K(t)$. The kernel $K(t)$ thus
decay faster than the power law, and its time derivative $\dot{K}(t)$ at
asymptotical times is described by an exponential decay given by
Eq.~\eqref{realtimeK}.

To account for the dynamics of $K(t)$ at all times, we need the knowledge of all solutions to the relevant transcendental equation
$\omega^2-\omega_0^2-\operatorname{Re}\Sigma_{x}(\omega)-i\operatorname{Im}\Sigma_{x}(\omega)=0$. It is highly improbable to analytically locate all of them; nonetheless, when $|\omega|$ is much larger than unity, with the help of Eqs.~\eqref{resigma-x} and \eqref{imsigma-x}, this equation may be transformed into a exponential polynomial such that the
asymptotic solutions can be found. We decompose the solution $\omega$ into its real part $u$ and imaginary part $v$, i.e. $\omega=u+iv$.
The
coupled equations of $u$ and $v$ can be given by
\begin{align*}
    \ln\sqrt{u^{2}+v^{2}}-2z_{0}v&=\ln\frac{3}{4 z_0}\,,\\
    \tan^{-1}\frac{v}{u}+2z_{0}u&=(2n+1/2)\pi\,,\qquad\qquad  n \in\mathbb{Z} \,.
\end{align*}
Thus asymptotically, the solutions are evenly distributed in the $u$
sense along the curve $\ln\sqrt{u^{2}+v^{2}}-2z_{0}v=\ln( 3/4 z_0)$ on the
$u-v$ plane as shown in Fig.~\ref{Fi:branch}. With $|u|\gg|v|$, both
$u$ and $v$ are given by
\begin{align*}
    u&=\frac{1}{T}(2n+1/2)\pi\,,\qquad\qquad|n|\gg1\,,\\
    v&=\frac{1}{T}\left(-\ln \frac{3}{ 4 z_0}+\ln|u|\right)>0\,.
\end{align*}
It can be seen that $v$ is always positive so that all the
asymptotical solutions are  on the upper half side of the
complex $\omega$ plane. These modes will contribute to the motion
via the time evolution factor
\begin{equation}
    \exp(i\omega t)\propto n^{-\frac{t}{T}}\exp(i\,2n\pi\frac{t}{T})\,,\qquad\qquad|n|\gg1\,.
\end{equation}
Their contributions are thus negligible for the dynamics of
$\dot{K}_{x}(t)$ at times $t\gg T$. These relatively high frequency
modes are significant only in the very early stage of the
evolution. Therefore, the resonance mode, which peaks at
$\Omega$ with width $\Gamma$, have the dominant effect for the late-time dynamics of the non-Markovian Langevin equation in a weak coupling limit. The memory effect or the effect
of the time-delay will register in the parameters such as the
resonance frequency $\Omega$, the decay constant $\Gamma$, and so
on.

The Laplace transform method seems too complicated to properly take
all asymptotical resonance modes into consideration in order to deal
with the very early-time behavior of the charged oscillator. It
would be more straightforward to apply the iteration techniques. The
idea goes as follows. From the initial conditions in the interval
$-T<t<0$, we may find the solution to Eq.~\eqref{time-delay-xx} for
the times $0<t<T$. Because in this time interval, the retardation
effect does not settle in yet, the equation of motion essentially
describes the Markovian motion due to the damped oscillator with a
driving source. Next, this solution is then plugged into the
retardation terms on the right hand side of
Eq.~\eqref{time-delay-xx} to iteratively find the solution for the
time interval $T<t<2T$. We may continue with these procedures, but
the secular terms are expected to appear after several
iterations. The method of dynamical renormalization group can then
be implemented by resumming these secular terms to address more on
the non-Markovian nature of the full-time dynamics of the
solution~\cite{DB}.

\subsection{perpendicular motion}
We now consider a charged particle moving in the $z$ direction
perpendicular to the plane boundary. Contrary to the previous case,
the modified instantaneous Coulomb term due to the boundary will
make a shift to the oscillation frequency of the motion. Besides, it
causes a static attraction between the charge and its image, so an
applied potential is introduced to counteract the attractive Coulomb
force. The corresponding linearized Langevin equation for the
displacement from the equilibrium position is obtained from Eqs.~\eqref{nonlinearlangevin}, \eqref{ret}, and \eqref{Hadamrd} as
\begin{align}\label{zzz}
     m\ddot{q}^{z}(t)+\left[\partial_{z}^{2}V(z_0)-\frac{e^2}{4\pi}\frac{4}{(2z_0)^3}\right]q^{z}(t)+\frac{e^{2}}{4\pi}\frac{1}{(2z_{0})^{2}}&+e^2\int_0^tdt'\;\dot{g}_{R}^{z}[z_0,z_0;t-t']\,\dot{q}^{z}(t')\notag\\
     &=-\,\hbar\,e\,\dot{\xi}^z(t)\,,
\end{align}
where the $zz$ component retarded Green's function $g_{R}^{z}$ can be expressed in terms of the spectral density,
\begin{equation*}
     {g}_{R}^{z}[z_0,z_0;t-t']=-\theta(t-t')\int_0^{\infty}\frac{dk}{\pi}\;\rho^{z}(z_0,z_0;k)\,\sin[k(t-t')]\,,
\end{equation*}
with
\begin{equation}
    \rho^{z}(z_0,z_0;\omega)=-\frac{k}{\pi}\left[\frac{1}{3}-\frac{\cos(2k z_0)}{(2k z_0)^2}+\frac{\sin(2k z_0)}{(2k z_0)^3}\right]\,.\label{spectralden-zz}
\end{equation}
The noise-noise correlation functions are then obtained as
\begin{align}
     \langle\xi^{z}(t)\rangle&=0\,,\qquad\qquad\langle\xi^{z}(t)\xi^{z}(t')\rangle=\frac{1}{\hbar}\,g_{H}^{z}[z_0,z_0;t-t']\,,\\
     g_{H}^{z}[z_0,z_0;t-t']&=-\int_0^{\infty}\frac{dk}{2\pi}\;\rho^{z}(z_0,z_0;k)\,\cos[k(t-t')]\,.\label{hadama-z}
\end{align}
As discussed before, a fluctuation-dissipation relation is obeyed.

After mass renormalization, a similar time-delay differential
equation for the charged oscillator moving in the $z$ direction in the presence of the boundary is given by
\begin{align}
 m_{r}\ddot{q}^{z}(t)+\left[m_{r}\omega^{2}_0-\frac{e^2}{4\pi}\frac{2}{(2z_0)^3}\right]q^{z}(t)= -\,\hbar\,e\,\dot{\xi}^z(t)&+\frac{e^2}{6\pi}\dddot{q}^{z}(t)\notag\\
    &+\frac{2e^{2}}{4\pi}\left[\frac{\dot{q}^{z}(t-T)}{(2z_0)^2}+\frac{q^{z}(t-T)}{(2z_{0})^3}\right]\,.
\end{align}
The applied potential $V(\mathbf{q})$ has been turned on to drive
the charged particle into a harmonic motion of frequency
$\omega_{0}$ in the $z$ direction. In addition, the retarded Green's
function also has the contribution to the frequency shift. This
equation explicitly shows the memory effect due to the boundary. In
particular, all backreaction terms arising from the boundary come
from the velocity field. The contribution of the acceleration field
vanishes under the dipole approximation since the relative direction
from the image charge to the charge itself is perpendicular to its
motion. Then, the real-time solution $q^{z}$ to the inhomogeneous
stochastic equation~\eqref{zzz} takes a similar form to
Eq.~\eqref{realtimesol}. The time evolution of the kernel $K_{z}(t)$
from the inverse Laplace transform is obtained as
\begin{equation}\label{Kz}
    \dot{K}_z(t)\sim Z_z\,\cos(\Omega_{z}t+\alpha_z)\,e^{-\Gamma_z t}\,,\qquad t>0\,,
\end{equation}
with the resonance frequency $\Omega_{z}$ and the decay constant
$\Gamma_z$ given by
\begin{equation}\label{E:sd}
\Omega_{z}\sim  \omega^{2}_0 -\frac{e^2}{4\pi}\frac{4}{m_{r}(2z_0)^3}  +\frac{\operatorname{Re}\Sigma_{z}(\omega_0)}{2\omega_0}\,,\quad\quad \Gamma_z\sim Z_z\frac{\operatorname{Im}\Sigma_{z}(\Omega_z)}{2\Omega_z}\,,
\end{equation}
respectively where the frequency shift due to the boundary corrections of the instantaneous Coulomb potential has been taken into account.
Moreover, the phase shift $\alpha_z$ and $Z_z$ are found to be
\begin{equation}
     Z_z\sim\left[1-\frac{\partial\operatorname{Re}\Sigma_{z}(\Omega_z)}{\partial\Omega_z^2}\right]^{-1}\,,\quad\quad\alpha_z\sim Z_z\frac{\partial\operatorname{Im}\Sigma_{z}(\Omega_z)}{\partial\Omega_z^2}\,. \label{rewavefun-phase-z}
\end{equation}
Both the mass and the real part of the $\Sigma_{z}(\omega)$ kernel
have been renormalized. The corresponding
$\widetilde{\Sigma}_{z}(s)$ kernel is
\begin{align}
    \widetilde{\Sigma}^{z}(s)&=\frac{e^2}{m}\,s^2\,\widetilde{g}_{R}^{z}[z_0,z_0;s]\notag\\
    &=-\frac{e^2}{m} \, \left[\int^{\Lambda}\frac{dk}{\pi}\;k\, \rho^{z}(z_0,z_0;k)\, \frac{s^2}{s^2-k^2}-\frac{\Lambda}{3\pi^2} \right] \, ,\label{resigma}
\end{align}
and the real and imaginary parts of $ \widetilde{\Sigma}_{z}(s)$ in the vicinity of the branch-cut are given, respectively, by
\begin{align}
    \operatorname{Re}\Sigma_{z}(\omega)&=\frac{e^2}{4\pi m}\,\omega^3\left[\frac{2}{(2\omega z_0)^3}-\frac{2\sin(2\omega z_0)}{(2\omega z_0)^2}-\frac{2\cos(2\omega z_0)}{(2\omega z_0)^3}\right]\,,\\
    \operatorname{Im}\Sigma_{z}(\omega)&=\frac{e^2}{4\pi m}\,\operatorname{sgn}(\omega)\,\omega^3\left[\frac{2}{3}-\frac{2\cos(2\omega z_0)}{(2\omega z_0)^2}+\frac{2\sin(2\omega z_0)}{(2\omega z_0)^3}\right]\,. \label{imsigma-z}
\end{align}
The presence of the boundary apparently modifies the behavior of the
charged oscillator in an anisotropic way. This characteristic is
especially noticeable near the boundary where the electric fields
parallel to the plate vanish, but their normal components become
doubled, compared with their counterparts without the boundary. This anisotropic feature is encoded in the spectral density. Thus, the
quantities that can be expressed with the spectral density should share the
same feature. As a result, the decay constant $\Gamma$ for
motion parallel to the boundary turns out to be smaller than in
the perpendicular case with a similar configuration. Additionally, the
effect of the stochastic noise on the oscillator ought to be much
weaker in the parallel case than the perpendicular one.

Next section will be devoted to studying the time evolution of velocity
fluctuations of the charged oscillator due to both anisotropically modified
vacuum fluctuations by the boundary and associated dissipative backreaction from electromagnetic fields.

\section{velocity fluctuations }\label{sec5}
When the charged particle couples to quantized electromagnetic
fields, the stochastic Lorentz force, manifested from vacuum fluctuation of fields, drives
the particle into a fluctuating trajectory in analogy to the Brownian
motion. Thus, it is of interest to study velocity fluctuations of
the charged oscillator to see how they are affected by the boundary
and are asymptotically saturated as a result of the
fluctuation-dissipation relation. We will compute them in the
interval $1/\Omega \ll t \ll 1/\Gamma $ for the linearly growing regime
in which backreaction dissipation can be ignored, as well as in the
interval of $1/\Gamma\ll t$ for the saturation regime where the effects of
fluctuations and dissipation come into balance.

As the particle starts to move at $t=0$, its velocity fluctuations at
time $t$ in the direction $i$ driven by electromagnetic vacuum
fluctuations  are given by
\begin{equation}\label{E:vel fluc}
   \langle\Delta v_i^2(t)\rangle= \frac{e^2}{m^2}\int_0^tdt'\int_0^t dt''\;\dot{K}_i(t')\frac{d^2}{dt'dt''}\Bigl[g_{H}^{i}(z_0,z_0;t'-t'')\Bigr]\dot{K}_i(t'')\,,
\end{equation}
where $g_{H}^{i}$ is the Hadamard function of vector potentials in
the dipole approximation.  We have implicitly assumed that
$\langle\Delta v_i^2\rangle$ vanishes initially. The charged
oscillator is allowed to move either parallel or perpendicular to
the boundary, and thus the relevant Hadamard function takes the form
of Eq.~\eqref{hadama-x} or \eqref{hadama-z}. The function $K_i(t)$
is the kernel of the equation of motion, and its time derivative is
approximated by Eq.~\eqref{realtimeK} or \eqref{Kz}. This
approximate solution holds for $t> 1/\Omega_{i} \ge 2 z_0$ by
ignoring the contributions from high-frequency modes.
Then, it may give rise to errorr in computing the integral \eqref{E:vel fluc}
for the time regime $t\le2z_0$. In fact, one can argue that the
error for a large distance $z_0$ can be trivially neglected because
the boundary correction is negligible. Moreover, although the
presence of the boundary will result in significant contributions
for small $z_0$, the time interval $0\le t\le 2 z_0$ over which
integrations in Eq.~\eqref{E:vel fluc} are performed, are also small.
Only a minor error is introduced.

In terms of the spectral density, the velocity fluctuations becomes
\begin{align}
     \langle\Delta{}v_i^2(t)\rangle&=-\frac{e^2}{m^2}\int_0^tdt'\int_0^tdt''\;\dot{K}_i(t')\dot{K}_i(t'')\int_{0}^{\infty}\frac{dk}{2\pi}\;k^{2}\rho^{i}(z_{0},z_{0};k)\cos[k(t'-t'')]\notag\\
             &=-\frac{e^2}{m^2}\int_{0}^{\infty}\frac{dk}{2\pi}\;\, k^{2}\, \rho^{i}(z_{0},z_{0};k)\, I(t;k)\,,\label{E:vel_flu}
\end{align}
where
\begin{equation}
    I(t;k)=\int_0^tdt'\int_0^t dt''\;\dot{K}_i(t')\dot{K}_i(t'')\cos[k(t'-t'')]\,.
\end{equation}
In writing so, the integrand of the $k$-integral is factorized into
a product of the spectral density $\rho^{i}(z_{0},z_{0};k)$ and the
function $I(t;k)$, which solely depend on the distance $z_{0}$ and
time $t$, respectively. The full expression of the function $I$ is
too intricate to be of any help, but in the time intervals
$1/\Omega_{i} \ll t \ll 1/\Gamma_i$ and $t \gg 1/\Gamma_i $, it can be
reduced greatly,
\begin{enumerate}
     \item $1/\Omega_{i} \ll t \ll  1/\Gamma_{i}$: linearly growing regime, with $\Gamma_{i}\,t$ set to zero,
            \begin{align*}
                I_0(t, k)&\equiv I(\frac{1}{\Omega_{i}}\ll t\ll\frac{1}{\Gamma_i};k)\\
                &=\frac{Z_i^2}{2(k^2-\Omega_{i}^2)^2}\bigg\{2(k^2+\Omega_{i}^2)+(k^2-\Omega_{i}^2)\cos(2\alpha_i)+(k^2-\Omega_{i}^2)\cos(2\Omega_{i} t+2\alpha_i)\biggl.\\
                             &\qquad\qquad\qquad\qquad-(k+\Omega_{i})^2\cos[(k-\Omega_{i})t]-(k-\Omega_{i})^2\cos[(k+\Omega_{i})t]\\
                             &\biggr.\qquad\qquad-(k^2-\Omega_{i}^2)\cos(\Omega_{i} t-k  t+2\alpha_i)-(k^2-\Omega_{i}^2)\cos(\Omega_{i} t+ k t+2\alpha_i)\biggr\}\,.
            \end{align*}
    \item $\Gamma_i \,t\gg1$: saturation regime, with $\Gamma_i\,t$ set to infinity,
            \begin{equation*}
                I_{\infty}(k)\equiv I(t\to\infty;k)=Z_i^2\frac{\left(k^2+\Gamma_i^2\right)\cos^2\alpha_i-2\Omega_{i}\Gamma_i\cos\alpha_i\sin\alpha_i+\Omega_{i}^2\sin^2\alpha_i}{\left[\left(k-\Omega_{i}\right)^2+\Gamma_i^2\right]\left[\left(k+\Omega_{i}\right)^2+\Gamma_i^2\right]}\,.
            \end{equation*}
\end{enumerate}
Notice that the function $I$ has a Breit-Wigner feature in the $k$
space with the resonance peak at about $\Omega_{i}$ and its width
being approximately of order $\pi/t$ at early times or $\Gamma_i$
for the late-time regime. The narrow width may result in a prominent
peak. On the other hand, the spectral density $\rho^{i}$ reveals
the oscillatory behavior on the scales $1/z_0$ in the $k$ space, which will lead to heavy cancelation as
the distance $z_{0}$ is sufficiently large so that the boundary
effect becomes insignificant. Thus, the result of the integration
\eqref{E:vel_flu} relies on two competing scales $1/z_{0}$, and
$\pi/t$ or $\Gamma_i$. Additionally,  the behavior of the integrand
in Eq.~\eqref{E:vel_flu} increases linear in $k$ when $k$ is
sufficiently large. Therefore, velocity fluctuations after doing
integration over $k$ will have a quadratic divergence, which has to
be regularized by introducing a cutoff. $Z_{i}$ is
approximately equal to unity in the weak coupling limit.

\subsection{Growing regime}
Here we study the motion of the charged oscillator at the early stage for
$1/\Omega_{i}\ll t\ll1/\Gamma_i$ when dissipation backreaction can be
ignored. Velocity fluctuations thus mainly result from the
stochastic noise. For $t\gg2z_{0}$ when the retardation effect out
of the boundary becomes effective, we may expect that the spectral
density varies relatively slowly with $k$ in comparison with the function $I$
around the resonance peak.

Therefore, the integration \eqref{E:vel_flu} can be approximated by
pulling the spectral density $\rho^{i}$ out of the integral and
evaluating it with $k\sim \Omega_{i}$ in the neighborhood of the
peak. With this in mind, we then rewrite Eq.~\eqref{E:vel_flu} in
terms of the dimensionless parameters denoted as $\gamma_i
=\Gamma_i/\Omega_{i}$, $\bar{z}_{0 i}=\Omega_{i}z_{0}$, $y=k/\Omega_{i}$,
and $\tau_i=\Omega_{i}t$, and obtain
\begin{equation}\label{E:dnsedU}
     \langle\Delta v_i^2(t)\rangle= -\frac{e^2\Omega_{i}^{2}}{m^2}\int_{0}^{\infty}\frac{dy}{2\pi}\;y^{2}\bar{\rho}^{i}(\bar{z}_{0 i},\bar{z}_{0 i};y)\,\mathcal{I}(\tau_i;y)\,,
\end{equation}
where
\begin{equation*}
     \bar{\rho}^{i}(\bar{z}_{0 i},\bar{z}_{0 i};y)=\frac{1}{\Omega_{i}}\rho^{i}(z_{0},z_{0};k)\,,\qquad\qquad\mathcal{I}(\tau_i;y)=\Omega_{i}^{2}\,I_{0}(t;k)\,.
\end{equation*}
Since the function $\mathcal{I}$ demonstrates a sharp peak about $y=1$ with
the width of order $\pi/\tau_i$, as long as $\bar{z}_{0 i}/\tau_i \ll1$,
the velocity fluctuations can be approximately given by
\begin{equation}
     \langle\Delta v_i^2(t)\rangle  \sim \langle\Delta v_i^2(t)\rangle_{\text{div.}}
     -\frac{e^2\Omega_{i}^{2}}{m^2}\,\bar{\rho}^{i}(\bar{z}_{0 i},\bar{z}_{0 i};1)\int_{0}^{\infty}\frac{dy}{2\pi}\;\mathcal{I}(\tau_i;y) \, , \label{v;growing}
\end{equation}
where
\begin{equation}
    \int_{0}^{\infty}\frac{dy}{2\pi}\;\mathcal{I}(\tau_{i};y)\approx\frac{\tau_{i}}{4}+\text{terms oscillatory with time}\,
\end{equation}
leads to a contribution linearly growing in time. The high-frequency contributions $\langle\Delta v_i^2(t)\rangle_{\text{div.}}$ need
regularization by inserting a convergent factor $e^{-y\epsilon_i}$
into the integrand. Here $\epsilon_i $ is the dimensionless
short-distance cutoff. Then, it ends up with
\begin{align}
     \langle\Delta v_i^2(t)\rangle&\sim \frac{\Omega_{i}}{m}\,\Gamma_{i}\,t+\frac{e^2\Omega_{i}^{2}}{16\pi^2m^2}\,\left.\Bigl[2+\cos(2\alpha_i)+\cos(2\Omega_{i}t+2\alpha_i)\Bigr]
     \begin{cases}
        \frac{4}{3 \epsilon^2_i}\,,&\text{$x$ direction}\,,\\
        \frac{4}{3 \epsilon^2_i} + \frac{1}{\bar{z}_{0 i}^2}\,,&\text{$z$ direction}\,,
     \end{cases}\right\}\notag\\
     &\sim \frac{\Omega_0}{m}\,\Gamma_{i}\,t+\frac{e^2}{16 \pi^2 m^2}\,\left.\Bigl[3+\cos(2 \,\Omega_0\, t)\Bigr]
     \begin{cases}
        \frac{4}{3 \lambda^2_{\text{dB}}}\,,&\text{$x$ direction}\,,\\
        \frac{4}{3 \lambda^2_{\text{dB}}} + \frac{1}{z_0^2}\,,&\text{$z$ direction}\,,
    \end{cases} \right\}\,.
\end{align}
Notice that to obtain the last expression,
the dimensionless short-distance cutoff $\epsilon_{i}$
is chosen as a product of the resonance frequency
$\Omega_{i}$ and the de Broglie wavelength of the charge
$\lambda_{\text{dB}}$, that is,
$\epsilon_{i}=\Omega_{i}\lambda_{\text{dB}}$. The cutoff
dependence terms are known to arise from the unbounded Minkowski vacuum fluctuations, experienced by the charged oscillator. They weakly depend on the  modified oscillatory motion that results from the coupling to vacuum fluctuations in the presence of boundary. We will make further approximations by $\Omega_i \sim \Omega_0$, $\alpha_i \sim 0,$ and $ Z_i \sim 1 $, ignoring higher order effect on $z_0$. For motion perpendicular to the boundary, the dominant $z_0$ dependence of high frequency contributions is thus the $1/z_0^2$ term, and should be small as
compared with the cut-off dependence term in accordance with the classical assumption on the particle. However, the corresponding term is missing for motion parallel to the boundary. This can be understood as the consequence that there exists no corresponding boundary for the motion in the $x$ direction; thus no length scale is introduced in this direction.

Now we consider the ratio of the cut-off dependence term to the term linear in time, obtained form the above expression. With the help of the explicit expressions of the relaxation constants $\Gamma_i$, given by Eqs.~\eqref{imsigma-x}, \eqref{frequency-decay}, \eqref{E:sd}, and \eqref{imsigma-z},
\begin{eqnarray}
\Gamma_x (  \Omega_0) &=& \frac{e^2}{8\pi m}\, \Omega_0^2\left[\frac{2}{3}-\frac{\sin(2 \Omega_0 z_{0})}{(2  \Omega_0 z_{0})}-\frac{\cos(2 \Omega_0 z_{0})}{(2 \Omega_0 z_{0})^{2}}+\frac{\sin(2 \Omega_0 z_{0})}{(2  \Omega_0 z_{0})^{3}}\right]\,,\label{decat}\\
\Gamma_z (  \Omega_0) &=& \frac{e^2}{8\pi m}\, \Omega_0^2\left[\frac{2}{3}-\frac{2\cos(2  \Omega_0 z_0)}{(2 \Omega_0 z_0)^2}+\frac{2\sin(2 \Omega_0 z_0)}{(2 \Omega_0 z_0)^3}\right] \,, \label{decay}
\end{eqnarray}
it can be shown that the ratio is of the order
\begin{equation*}
   \mathcal{O}(\frac{1}{\Omega_0 t}\,\frac{1}{\epsilon^2})\sim \mathcal{O}(\frac{1}{\Omega_0 t}\,\frac{1}{ \Omega_0^2 \lambda_{dB}^{2}})\,.
\end{equation*}
Apparently, the high-frequency contributions can be dominant for the
sufficiently low oscillation frequency, or at very early stage of
motion~\cite{YU}. At later times, $\epsilon^{-2}\Omega_0{}^{-1}\ll
t$, they then become insignificant as it can be further seen from the numerical result to be discussed later

Therefore, the velocity fluctuations are found to grow linearly with time.
The growing rate is related to the relaxation constant $\Gamma_i$, evaluated at the resonance frequency $\Omega_0$. Backreaction dissipation will set in at later times, and will asymptotically counteract the
effects of fluctuations.

\subsection{Saturation regime}
Next we will investigate the behavior of velocity fluctuations at
much later times $t\gg 1/\Gamma_i$ by incorporating backreaction
dissipation. In this regime, they are described by
\begin{equation}\label{E:dnsedF}
     \langle\Delta v_i^2\rangle=-\frac{e^2\Omega_{i}^{2}}{m^2}\int_{0}^{\infty}\frac{dy}{2\pi}\;y^{2}\bar{\rho}^{i}(\bar{z}_{0 i},\bar{z}_{0 i};y)\,\mathcal{I}(y)\,,
\end{equation}
where
\begin{equation*}
     \mathcal{I}(y)=\Omega_{i}^{2}\,I_{\infty}(k)\,.
\end{equation*}
In general, since Eq.~\eqref{E:dnsedF} has a quadratic divergence, a
regulator of the form $e^{-y\epsilon_{i}}$ will be introduced to
regularize the integral. The influence from the boundary can be
significant when the dimensionless distance of the oscillator
from the boundary $\bar{z}_{0 i}$ is much smaller than unity.

Thus, we will expand the result of velocity fluctuations in terms of
small $\bar{z}_{0 i}$, after we plug in the $\bar{z}_{0 i}$ dependence of the
parameters $\gamma_{i}$, $\alpha_{i}$, and $\bar{\rho}^i$ for the
motion in the direction $i$, which are given by,
\begin{align*}
    &\bar{\rho}^{x}(\bar{z}_{0 x},\bar{z}_{0 x}; y) =-
     \frac{y}{\pi}\left[\frac{1}{3}-\frac{\sin(2\bar{z}_{0 x})}{2(2\bar{z}_{0 x})}-\frac{\cos(2\bar{z}_{0 x})}{2(2\bar{z}_{0 x})^{2}}+\frac{\sin(2\bar{z}_{0 x})}{2(2\bar{z}_{0 x})^{3}}\right]\,,\\
    &\bar{\rho}^{z} ( \bar{z}_{0 z}, \bar{z}_{0 z}; y)=- \frac{y}{\pi}\left[\frac{1}{3}-\frac{\cos(2\bar{z}_{0 z})}{(2\bar{z}_{0 z})^{2}}+\frac{\sin(2\bar{z}_{0 z})}{(2\bar{z}_{0 z})^{3}}\right]\,,\\
    &\alpha_{x}=\bar{r}_{c  x}\left[1-\frac{\cos(2\bar{z}_{0 x})}{2}+\frac{\sin(2\bar{z}_{0 x})}{2(2\bar{z}_{0 x})}\right]\,,\qquad\qquad\alpha_{z}=\bar{r}_{c  z}\left[1+\frac{\sin(2\bar{z}_{0 z})}{(2\bar{z}_{0 z})}\right]\,,
\end{align*}
and the dimensionless relaxation constant can be obtained from Eqs.~\eqref{decat} and \eqref{decay} as
$\gamma_i=-\pi\,\bar{r}_{c  i}\,\bar{\rho}^i$ with
$\bar{r}_{c  i}=\Omega_{i}\,r_{c}$. The parameter $r_{c}=e^2/(4\pi m)$ is the particle's
classical radius. Here, another small parameter $\bar{r}_{c  i}$,
characterizing the distance $r_c$ over the oscillation time scales due to a nonrelativistic motion, will also be used to extract the dominant contributions.

The saturated value of  velocity fluctuations can be found
analytically. For the motion parallel to the boundary, it is given
by
\begin{align}\label{E:vx}
     \langle\Delta v_x^2\rangle&=Z_x^2\frac{e^2}{16\pi^2m^2}\,\Omega_{x}^2\left\{\frac{8}{3\epsilon_{x}^{2}}+\left[\frac{2\pi}{\bar{r}_{c  x}}+\frac{16}{3}\ln\frac{2\bar{z}_{0 x}}{\epsilon_{x}}-\frac{40}{3}\right]\right.\notag\\
         &\qquad\qquad\qquad+\left.\left[\frac{776}{75}-\frac{32}{5}\gamma_{\epsilon}-\frac{32}{5}\ln2\bar{z}_{0 x}\right]\bar{z}_{0 x}^{2}+\mathcal{O}(\bar{z}_{0 x}^{4}\ln\bar{z}_{0 x}\,,\bar{r}_{c  x})\right\}\,\notag
         \\
         &\sim\frac{e^2}{16\pi^2m^2}\,\Omega_0^2\,\biggl\{\frac{8}{3\epsilon^{2}}+\frac{2\pi}{\bar{r}_c}\biggr\}=\frac{e^2}{8 \pi^2m^2} \,\biggl\{\frac{4}{3\lambda_{dB}^{2}}+\frac{\pi \, \Omega_0}{r_{c}}\biggr\}\,.
\end{align}
For the perpendicular case, we find
\begin{align}\label{E:vz}
     \langle\Delta v_z^2\rangle&=Z_z^2\frac{e^2}{16\pi^2 m^2}\,\Omega_{z}^2 \left\{\frac{8}{3\epsilon_{z}^{2}}+\frac{2}{\bar{z}_{0 z}^{2}}+\left[\frac{2\pi}{\bar{r}_{c  z}}-\frac{16}{3}\biggl(\ln2\bar{z}_{0 z}+\ln\epsilon_{z}\biggr)+\frac{40}{3}-\frac{32}{3}\,\gamma_{\epsilon}\right]+\right. \notag \\
         &\qquad\qquad\qquad+\left.\left[-\frac{448}{75}+\frac{16}{5}\gamma_{\epsilon}+\frac{16}{5}\ln2\bar{z}_{0 z}\right]\bar{z}_{0 z}^{2}+\mathcal{O}(\bar{z}_{0 z}^{4}\ln\bar{z}_{0 z}\,,\bar{r}_{c  z})\right\}\,\notag\\
         &\sim\frac{e^2}{16\pi^2 m^2}\,\Omega_0^2\,\biggl\{\frac{8}{3\epsilon^{2}}+\frac{2}{\bar{z}_0^{2}}+\frac{2\pi}{\bar{r}_{c}}\biggr\}=\frac{e^2}{8\pi^2m^2}\,\biggl\{\frac{4}{3\lambda_{dB}^{2}}+\frac{1}{z_{0}^{2}}+\frac{\pi \, \Omega_0}{r_{c}}\biggr\}\,,
\end{align}
where $\gamma_{\epsilon}$ is Euler's constant, with numerical value $\sim0.577216$.
The last lines of Eqs~\eqref{E:vx} and~\eqref{E:vz} are further approximated by
$Z_i\sim 1$, $\Omega_{i}\sim\Omega_0$, ignoring
higher-order $\bar{z}_0$-dependent terms where
$\bar{z}_0=\Omega_0 z_0$, $\bar{r}_{c}=\Omega_0
r_c$, and $\epsilon=\Omega_0\lambda_{dB}$; hence typically $\bar{z}_0>\epsilon\gg\bar{r}_{c}$. The parameter $\Omega_0$ is the renormalized oscillation frequency with the frequency shift due to the interaction with
fields. Here for simplicity, the same $\Omega_0$ is
chosen for the motion in both directions because the anisotropy it
introduces is of next order in $e^2$. In general, the results
\eqref{E:vx} and \eqref{E:vz} should show strong dependence on the relative
orientation between the boundary and the direction of the motion for small $z_{0}$,
and thus are anisotropic. It can be understood as a result of vacuum
fluctuations of electromagnetic fields under the influence of the
boundary~\cite{JT}. The enhancement in velocity fluctuations due to the presence of the boundary in the direction normal to the conducting plate arises from large induced $\mathbf{E}$-field fluctuations near the boundary, in comparison with the fluctuations parallel to
the plate~\cite{YU}. On the other hand, the term depending on $\Omega_0$ results from particle's motion, and is found to be boundary-independent. The physics behind this features can be explored in a more general context as
follows. Starting from Eq.~\eqref{E:dnsedF}, for small $\bar{z}_{0 i}$, and
$\bar{r}_{c  i}$, the contribution from the resonance peak is given by
approximating the integral over $y$ with the spectral function
evaluated at the dimensionless resonance frequency, $y\sim 1$,
\begin{align*}
    \int_{1-\gamma}^{1+\gamma}\frac{dy}{2\pi}\;y^{2}\bar{\rho}^{i}(\bar{z}_{0 i},\bar{z}_{0 i};y)\,\mathcal{I}(y)&\sim\bar{\rho}^{i}(\bar{z}_{0 i},\bar{z}_{0 i};1)\int_{1-\gamma}^{1+\gamma}\frac{dy}{2\pi}\;\mathcal{I}(y)\\
         &\sim\bar{\rho}^{i}(\bar{z}_{0 i},\bar{z}_{0 i};1)\int_{0}^{\infty}\frac{dy}{2\pi}\;\mathcal{I}(y)\sim\frac{\bar{\rho}^{i}(\bar{z}_{0 i},\bar{z}_{0 i};1)}{8\gamma_{i}}\sim-\frac{1}{8\pi}\frac{1}{\bar{r}_{c}}\,,
\end{align*}
where again the approximation $\Omega_{i}\sim\Omega_0$
has been applied to obtain the last expression. Apparently, the
result superficially depends on the spectral function
$\bar{\rho}^{i}$; however, the integral to the right of the spectral
function yields an expression proportional to $1/\gamma_{i}$. Since the
dimensionless relaxation constant $\gamma_{i}$ can be related to the
spectral density by $-\pi\bar{r}_{c  i}\,\bar{\rho}^{i}(\bar{z}_{0 i},\bar{z}_{0 i};1)$,
the boundary dependence due to the spectral density $\bar{\rho}^{i}$ is
canceled. Thus, the result does not depend on the relaxation
constant, hence independent of the orientation. Recall that the function $\bar{\rho}^{i}$ is
proportional to the rate at which velocity fluctuations grow,
while the time scales to reach the saturated regime are given by
$1/\gamma_{i}$. Thus it implies that although the velocity fluctuations increases at different rates at early times, they  will take different amount of time to reach saturation. In the end, the saturated values
are independent of the direction of the motion, and are only determined by the
oscillation frequency of the charged particle. This cancelation of
the boundary dependence, in fact, can be understood as reminiscence
of the fluctuations-dissipation relation since the function
$\bar{\rho}^{i}$ comes from the noise kernel $g_{H}^{i}$  while the
dimensionless relaxation constant $\gamma_{i}$ is inherited from the
dissipation kernel $g_{R}^{i}$.

As stated above, the integrand of Eq.~\eqref{E:dnsedF} grows
linearly for large values of $y$, so the contribution of this
portion is given by
\begin{equation}\label{VF}
     \int_{\sigma}^{\infty}\frac{dy}{2\pi}\;y^{2}\bar{\rho}^{i}(\bar{z}_{0 i},\bar{z}_{0 i};y)\,\mathcal{I}(y)e^{-y\epsilon_{i}}\sim
     -\frac{1}{ 16 \pi^2}
            \begin{cases}
                    \dfrac{8}{3\epsilon^{2}}\,,&\text{in the $x$ direction}\,,\\
                    \dfrac{8}{3\epsilon^{2}}+\dfrac{2}{\bar{z}_0^{2}}\,,&\text{in the $z$ direction}\,,
             \end{cases}
\end{equation}
where $\epsilon=\Omega_0\lambda_{dB}$, and the lower limit $\sigma$ of the integral is chosen to be greater than unity but otherwise arbitrary in order to exclude the
integration region around the resonance peak. As long as $\bar{z}_{0 i}\sigma\ll1$, the
integral is pretty much independent of the choice of $\sigma$.
Clearly, the integral~\eqref{VF} is  boundary-dependent and thus
anisotropic. The quadratic dependence on the cutoff can be
characterized by the width of the particle's wave wavefunction. When
the spatial extension of the wave function is smaller, it tends to
probe the finer structure of  fluctuations, which in turn causes
larger variation in particle's velocity.

Some comments are in order. In the limit $\Omega_0\to
0$, the motion of the charged particle becomes  insignificant and
the dissipation effect is ignored. Hence, it is of interest to
compare our results with the earlier study~\cite{YU}. They are consistent as $t\to\infty$, apart from a factor
$1/2$ due to an average over a period in the oscillatory motion. In this limit, the backreaction dissipation
can be safely ignored because the relaxation constant $\Gamma_{i}$ with
$\Omega_0^{2}$ dependence, is vanishingly small for
such a slow motion. Nonetheless, for a finite value of
$\Omega_0$, the nonuniform motion results in a
dissipative effect,  which damps out the particle's motion.

\subsection{Numerical results}
\begin{figure}
\centering
    \scalebox{1}{\includegraphics{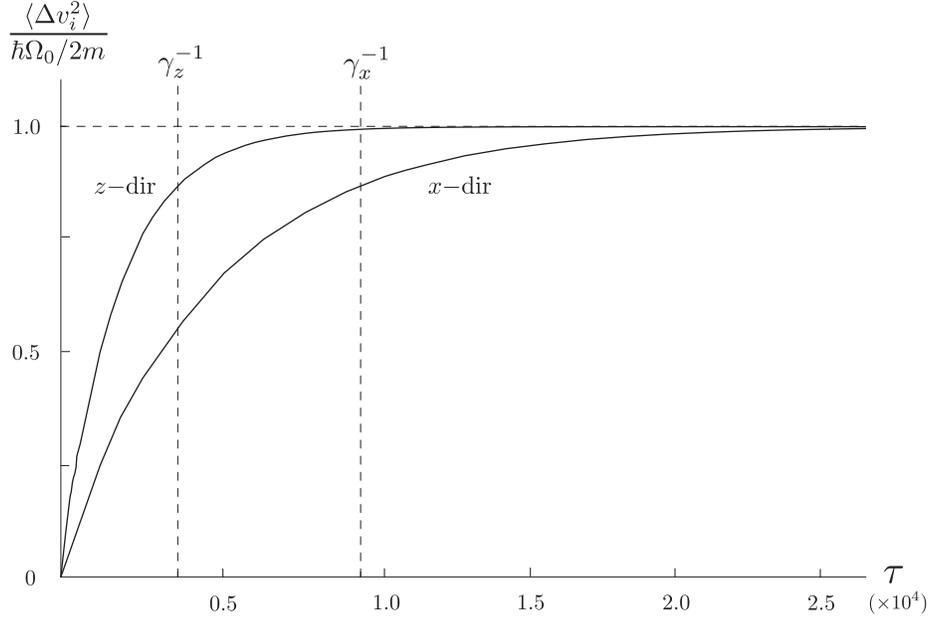}}
    \caption{Velocity fluctuations of the charge oscillator moving in the direction either parallel or normal to the plane boundary grow linearly at early times, and then saturate to a constant at late times. They start off at different rates, but approach the same saturated value $\hbar \Omega_0/2m$. The values of parameters are chosen as $\bar{z}_{0}=z_0 \Omega_0/c=1.0$ and $\bar{r}_{c}= r_c \Omega_0 /c =5\times10^{-4}$, and $\epsilon=0.1$; thus $\gamma_{z}^{-1}=0.363\times10^{4}$ and $\gamma_{x}^{-1}=0.931\times10^{4}$.} \label{Fi:velotime}
\end{figure}

The typical behavior of time variation of velocity fluctuations is
shown in Fig.~\ref{Fi:velotime}, where the results are generated
numerically. The value of the cutoff is set by the de Broglie wavelength,
$\lambda_{\text{dB}} \sim h/(m v)$, so the ratio of the charge's classical radius to its de Broglie wavelength is ${r}_c / \lambda_{\text{dB}} \sim (e^2/ h c) \,(v/c) \sim 10^{-4}$ with $v/c \sim 10^{-2}$ for nonrelativistic motion. In Fig.~\ref{Fi:velotime}, we choose $\bar{z}_0 = z_0 \Omega_0 /c \sim 1$, for example, when the distance to the plate $z_{0}$ is of the order $\mu\mathrm{m}$, the oscillation frequency is given by $\Omega_0 \sim 10^{14}\,\mathrm{s}^{-1}$ below the plasma frequency of the plate $10^{16}\,\mathrm{s}^{-1}$~\cite{JD}.
Now for an arbitrary charged particle, without loss of generality, we may take appropriate values of the parameters to numerically compute full evolution of velocity fluctuations. Here we let $m \sim 10^{3} m_e$ where $m_e$ is the mass of an electron, leading to the value of  $\epsilon= \lambda_{\text{dB}} \Omega_0/c \sim 0.1 $ and
$\bar{r}_{c}=r_{c} \Omega_0/c = 5\times10^{-4}$.
The horizontal axis is the dimensionless time
$\tau=\Omega_0\,t$, and the vertical axis is velocity
fluctuations, normalized by the parameter
$\hbar \Omega_0/ 2m$. It is seen that at early times, the velocity fluctuations increase linearly at different rates for motions parallel and perpendicular to the
conducting plate, and then saturate to the same value at late times. In
this case the cutoff-dependent terms is negligible except for very early times $t \le \epsilon^{-2}\, \Omega_0^{-1}  \sim 10^2 \, \Omega_0^{-1}$, which is vanishingly short when compared with the relaxation time scales for motion in either direction, about the order of $1/\Gamma\sim 10^4 \, \Omega_0^{-1}$. Thus, their contribution can not possibly be seen in Fig.~\ref{Fi:velotime}.

\section{Discussions and Concluding remarks}\label{sec6}
The anisotropic behavior of velocity fluctuations can be observed at the
early stage when the velocity fluctuations grow linearly in time.
It becomes more significant for the small value of $ z_0 \Omega_0/c$, and can be estimated analytically from the ratio obtained from Eq.~\eqref{decay}
as
\begin{equation}
  \frac{\Delta v^2_{x}}{\Delta v^2_{z}}\bigg|_{\epsilon^{-2}\Omega_0{}^{-1}\ll t \ll \Gamma_i^{-1}} \sim
  \frac{\Gamma_x}{\Gamma_z} \sim 0.4 \, (z_0 \Omega_0/c)^2  \,,\qquad z_0 \Omega_0 /c \ll1\,. \label{v-ratio}
\end{equation}
However, the values of $z_0 \Omega_0 /c $ cannot be arbitrarily small, and are constrained by the underlying assumptions. To be consistent with the dipole approximation, the amplitude of the charged oscillator should be much smaller than its distance to the plate, and can be set to the order of $10^{-2} z_0$, for example.  Thus, the corresponding velocity $v$ of the charged oscillator is about $v/c \sim 10^{-2} ( z_0 \Omega_0/c)$. It can be argued that the charge's motion
cannot be too slow since it may give rise to a large
position uncertainty to jeopardize the assumption of a
point-like particle in the stochastic approach. It is of interest to take the electron as an example.
Because the de Broglie wavelength, characterizing the size of the electron wavefunction, is
$\lambda_{\text{dB}} \sim h/(m_ev)\sim10^{-12}\,(v/c)^{-1}\,\mathrm{m}$, when the velocity is overly small, the spatial extension of the electron wavefunction can be the same as or even  larger than the distance to the plate $z_0$. Then it contradicts to the earlier assumptions.
Nevertheless, the velocity of the electron
may still be chosen as small as $v/c \sim 10^{-4} $, consistent with  $\lambda_{\text{dB}}  \ll z_0 \sim \mu \rm m$, ending up with $ z_0 \Omega_0 /c \sim 10^{-2}$.
Thus, the oscillation frequency for such a slow motion is reduced to the value of $\Omega_0\sim 10^{12}\,\mathrm{s}^{-1}$.
It will lead to the relaxation constant $\Gamma \sim ( e^2 \Omega_0 /m_e  c)\Omega_0 \sim 10 \,\mathrm{s^{-1}}$ from Eq.~\eqref{decay}. In addition, the dimensionless cutoff is then given by $\epsilon = \lambda_{\text{dB}} \Omega_0 /c \sim 10^{-4}$.
Thus, even for an electron in rather slow motion, appreciable anisotropy $\Delta v^2_{x}/\Delta v^2_{z} \sim 10^{-4} $ can be found during the time regime $10^{-4}\,\mathrm{s} < t < 0.1\,\mathrm{s}$ in which the cutoff dependence effect can be safely ignored while the saturation has not been reached yet in both directions.

At later times, the velocity fluctuations of the charged oscillator near the boundary reach saturation, independent of the orientation of the boundary, as shown in Fig.~\ref{Fi:velotime}. The saturated value is approximately given by the ${\Delta v^2_{i}} (\infty) \sim  \hbar \Omega_0/ 2 m $ as consequence of its nonuniform motion. The contribution of the velocity fluctuations, resulting from the imposition of the boundary conditions on electromagnetic fields, are at most the same order of magnitude as the cutoff-dependent terms. Both of them can be argued to be ignored at late times. Finally, the effective temperature corresponding to saturated velocity fluctuations for such Brownian motion can be estimated by
\begin{equation}
    T_{eff}\sim\frac{\hbar\Omega_0}{k_{B}}\sim10\left(\frac{\Omega_0}{10^{12}\,\mathrm{s}^{-1}}\right)\,{\rm K}\,,
\end{equation}
where $k_B$ is the Boltzmann constant.

To conclude, we study the influence due to quantized
electromagnetic fields on the motion of a nonrelativistic charged particle near the conducting plate.  The nonlinear, non-Markovian
Langevin equation of the particle is derived with the method of Feynman-Vernon influence
functional, and it incorporates both dissipation backreaction on the
charge in the form of the retarded self-force as well as the stochastic noise,
manifested from  vacuum fluctuations of  quantized electromagnetic fields. The dipole
approximation, an appropriate approximation for a non-relativistic motion, is implemented to find the solution to the Langevin equation. We consider that
the charged particle undergoes a small-amplitude oscillation in the
direction either parallel or perpendicular to the boundary plane.
The noise-averaged trajectory of this charged oscillator is governed by the classical
dynamics. Furthermore, the evolution of the kernel $K_{i}(t)$, obtained by solving the Langevin equation, is found to be dominated by the narrow resonance in the weak coupling limit where the ratio of the decay width $\Gamma_{i}$ over the oscillation frequency $\Omega_{i}$ is much less than unity, $\Gamma_{i}/\Omega_{i} \ll 1$. Thus, the memory effects or the the effects of
the time-delay on backreaction terms due to the presence of the
boundary are just to modify the quantities such as $\Omega_{i}$,
$\Gamma_{i}$ and so on. Then, velocity fluctuations of the charged oscillator
driven by stochastic forces are found to grow linearly with time in the early
stage of the evolution at the rate given by the relaxation constant. It
 turns out to be smaller in the parallel case than in the
perpendicular one with a similar configuration, and reveals strong anisotropic behavior.
 They are then asymptotically saturated as a result of the fluctuation-dissipation
relation at rather different relaxation time scales. However, we find that
the same saturated value is obtained for motion in both directions at late times,
resulting from delicate balancing effects between dissipation backreaction and accompanying fluctuations, and thus the value is mainly
determined by its oscillatory motion. So, at late times the effects from boundary-modified vacuum fluctuations on the velocity dispersion of the charged particle can be hardly seen.

The dipole approximation amounts to linearizing the Langevin equation obtained above. However, beyond the dipole approximation, one expects to introduce the additional drift effects on the dynamics of the
particle from the trajectory-dependent Green's functions. In particular, it may give rise to the noise-induced-drift forces owing to the correlations of stochastic forces. This effect has been considered in the context of the fast moving particle~\cite{HU2}, and may have observational consequences.

\begin{acknowledgments}
We would like to thank Bei-Lok Hu, Larry H. Ford, and Ray J. Rivers for
stimulating discussions. This work was supported in part by the
National Science Council, R. O. C. under grant
NSC95-2112-M-259-011-MY2, and the National Center for Theoretical Science in Taiwan.
\end{acknowledgments}

\end{document}